# Ni- and Co-struvites: Revealing crystallization mechanisms and crystal engineering towards applicational use of transition metal phosphates

Stephanos Karafiludis[1*], Ana Guilherme Buzanich[1], Zdravko Kochovski[2], Ines Feldmann[1], Franziska Emmerling[1,3], and Tomasz M. Stawski[1**]

[1]Federal Institute for Materials Research and Testing, Richard-Willstatter-Straße 11, 12489 Berlin, Germany

[2]Department for Electrochemical Energy Storage, Helmholtz-Zentrum Berlin for Materials and Energy, Hahn-Meitner Platz 1, 14109 Berlin, Germany

[3]Department of Chemistry, Humboldt-Universität zu Berlin, Brook-Taylor-Straße 2, 12489 Berlin

Stephanos Karafiludis: https://orcid.org/0000-0002-7257-6311;
Ana Guilherme Buzanich: https://orcid.org/0000-0001-5543-9924;
Zdravko Kochovski: https://orcid.org/0000-0001-8375-0365 ;
Ines Feldmann: https://orcid.org/0000-0003-2650-9270 ;
Franziska Emmerling: https://orcid.org/0000-0001-8528-0301;
Tomasz M. Stawski: https://orcid.org/0000-0002-0881-5808;

Corresponding authors: **tomasz.stawski@bam.de; *stephanos.karafiludis@bam.de

**Abstract**

Industrial and agricultural waste streams (waste water, sludges, tailings etc.) which contain high concentrations of $NH_4^+$, $PO_4^{3-}$ and transition metals are environmentally harmful and toxic pollutants. At the same time phosphorous and transition metals constitute highly valuable resources. Typically, separate pathways have been considered to extract hazardous transition metals or phosphate, independently from each other. Investigations on the simultaneous removal of multiple components have been studied only to a limited extent. Here, we report the synthesis routes for Co- and Ni-struvites ($NH_4MPO_4 \cdot 6H_2O$, M = $Ni^{2+}$, $Co^{2+}$), which allow for P, ammonia and metal co-precipitation. By evaluating different reaction parameters, the phase and stability of transition metal struvites, as well as their crystal morphologies, and sizes could be optimized. Ni-struvite is stable in a wide reactant concentration range and at different metal/phosphorus (M/P) ratios, whereas Co-struvite only forms at low M/P ratios. Detailed investigations of the precipitation process using ex situ and in situ techniques provided insights into the crystallization mechanisms/crystal engineering of these materials. M-struvites crystallize via intermediate colloidal nanophases which subsequently aggregate and condense to final crystals after extended reaction times. However, the exact reaction kinetics of the formation of a final crystalline product varies significantly depending on the metal cation involved in the precipitation process: several seconds (Mg) to minutes (Ni) to hours (Co) . The achieved level of control over the morphology and size, makes precipitation of transition metal struvites a promising method for direct metal recovery and binding them in the form of valuable phosphate raw materials. Under this paradigm, the crystals can be potentially up-cycled as precursor powders for electrochemical or (electro)catalytic applications, which require transition metal phosphates (TMPs).

## Introduction

Recovery and recycling of critical and essential elements has crucial importance for maintaining sustainable use of raw materials. Phosphorus is an important, yet, limited natural resource. It is widely used in modern agriculture, mainly in fertilizers, but this element is not fully recyclable. In the mid-term future it could be potentially depleted as a raw material due to the high demand and rapidly declining natural phosphorite ore deposits [1, 2]. Thus, phosphate rocks are included in the European Union's list of critical raw materials (CRMs) [3]. At the same time, due to the incomplete P-cycle in nature [4], phosphorous also causes major problems to the environment such as eutrophication of aquatic systems and is effectively lost for use. Therefore, due to simultaneous scarcity and abundance, the phosphorus recovery from agricultural, industrial, mining, or urban wastewaters have been an important factor in sustaining our global consumption and preservation of the natural environment. In general terms, phosphorous recovery relies on precipitation of P-bearing mineral phases [5, 6]. Therefore, phosphorous recovery cannot be conducted without considering the presence and the "co-recovery" of metal counter-ions. Alkaline earth metals are ubiquitous, and as such are neither of any particular interest, nor do they constitute an environmental risk. In contrast, many d-block-elements are pollutants and/or valuable resources. These factors are especially important in the case of the controlled precipitation/crystallization of struvite from waste streams which is already a promising P-recovery route [7, 8]. Struvite is $NH_4MPO_4 \cdot 6H_2O$, where $M^{2+}$ is usually $Mg^{2+}$, but other divalent cations such as $Ni^{2+}$ and $Co^{2+}$ can potentially substitute $Mg^{2+}$ in the struvite structure due to similar ionic radii [7, 9-11]. These recovered M-struvites, with M = Ni and/or Co, could act as storage materials for transition metals that may be potentially up-cycled for industrial applications as raw materials for transition metal phosphate (electro)catalysts (TMPs) [12-15]. It has been recently shown that struvite crystals can be directly thermally decomposed to form mesoporosity and have high surface areas of > 200 $m^2/g$ (I.e. upper reported values for TMPs) [13, 16]. In this regard metal struvites constitute direct and low-cost precursors for mesostructured phosphate materials, which do not require any "external" templates to achieve mesoporosity during their synthesis.

Although, the synthesis of Mg-struvite is relatively well understood [8, 17] and used in practice [18], investigations of crystallization of transition metal struvites are scarce. Most of these studies focus on the incorporation of transition metals [19-22] or other components [23] majorly through adsorption rather than the complete substitution of magnesium in the struvite structure [19, 24-26]. The adsorption routes are important for cation removal e.g. from waste water, however the TMPs for applications typically require very high transition metal loads (~100% stoichiometric vs ppms). Although, transition metal struvites share their fundamental crystal structure with Mg-struvite, their crystallization pathways, growth kinetics, thermodynamic stability or transformations differ among each other and from for the parent compound, Mg-struvite [16, 27]. In transition metal phosphate recovery routes aimed at potential up-cycling, the holistic approach towards the crystallization routes is crucial. The goal is to obtain a given phase, but it should also exhibit engineered or at least predictable properties such as crystal size, shape, aspect ratio etc. Such crystal engineering of a desired chemical compound determines its performance and scalability in industrial applications. Firstly, the optimized reaction conditions lead to efficient M-struvite precipitation, and therefore to significantly improved phosphorous, Ni, and Co. Secondly, such "engineered" M-struvite crystals may be further directly used as tailor-made upcycled precursor materials for specific applications e.g. in (electro)catalysis [14, 15, 23, 28-30].

Here, we present the effect of physicochemical conditions (initial pH, different c(reactants)) of aqueous solutions on the formation, growth and morphology of Ni- and Co-struvites. We compared our findings with the parent Mg-struvite. We highlight the systematic differences in crystallization pathways of M-struvites (M = $Mg^{2+}$, $Ni^{2+}$ and $Co^{2+}$) at all stages, where the early-stage stability and speciation of the precursor phases is linked to the overall evolution of struvite crystals. We show that the formation in all metal systems involves intermediate colloidal nanophases, which subsequently aggregate and condense to final crystals although their reaction kinetics differ significantly from each other.

**Methods**

*Synthesis*

$(NH_4)_2HPO_4$ (DAP) (ChemSolute, 99%), $MgCl_2 \cdot 6H_2O$, $NiSO_4 \cdot 6H_2O$ (ChemSolute, 99%) and $CoSO_4 \cdot 7H_2O$ (Alfa Aesar, 98 %) were used to synthesize metal phosphates under different reaction conditions from 0.004-0.3 M for $NiSO_4$, $CoSO_4$, $MgCl_2$ and 0.02-0.1 M for DAP resulting in metal-to-phosphate ratios (M/P) between 0.2-3 for all systems. All detailed reaction conditions are listed in the SI (SI: Table S1). Our work was primarily focused on Ni and Co, but for the sake of comparison we also synthesized and investigated common struvite ($NH_4MgPO_4 \cdot 6H_2O$) at several selected concentrations (SI: Table S1). At first 1 M stock solutions of DAP and the metal salts were prepared. The aqueous solutions were prepared by dissolving the calculated amount of salt in double ionized water (> 18 MΩ·cm). In a simple "one pot" approach, typically 100 ml of DAP and 100 ml of metal-bearing solution were mixed on a magnetic stirrer (300 rpm) at room temperature (25°C). By using DAP with a fixed Nitrogen/phosphate (N/P) ratio of 2 the occurring precipitation reaction of M-struvite was favored according to the following mass-balance equation (eq. [1]):

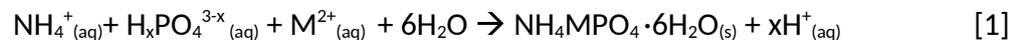

$$NH_4^+{}_{(aq)} + H_xPO_4^{3-x}{}_{(aq)} + M^{2+}{}_{(aq)} + 6H_2O \rightarrow NH_4MPO_4 \cdot 6H_2O_{(s)} + xH^+{}_{(aq)} \quad [1]$$

Eq. [1] was in general valid for Ni regardless of the metal concentration used, as well as for Co at lower concentrations below a M/P ratio of 0.4. However, at high concentrations of $Co^{2+}$ above a M/P ratio of 0.4 we observed the crystallization of cobalt(II)phosphate octahydrate (CPO) instead of Co-struvite (COS) according to eq. [2] (see later Figure 1):

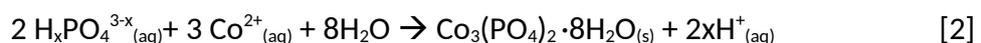

$$2\, H_xPO_4^{3-x}{}_{(aq)} + 3\, Co^{2+}{}_{(aq)} + 8H_2O \rightarrow Co_3(PO_4)_2 \cdot 8H_2O_{(s)} + 2xH^+{}_{(aq)} \quad [2]$$

To evaluate how the pH influences the crystal morphology, size and stability of M-struvites, sample series of Ni, Co and Mg were synthesized at different initial pH values by adding varying amounts (1-10 ml) of 1 M $NaOH_{(aq)}$ or 0.1 M $H_2SO_{4(aq)}$ into the glass reactor prior mixing of the reactants (DAP and metal bearing solution). Here, we used c($M^{2+}$) = 0.02 M and

c(DAP) = 0.1 M to remain for all chemical systems (Ni, Co and Mg) within the phase stability field of M-struvite (Figure 1). For further analysis, the solid phases precipitated from the mixed solutions were collected onto a cellulose filter (pore size 1 µm, LABSOLUTE) by using a vacuum filtration kit (i.e. Büchner funnel). Afterwards the powders were washed with distilled water and left to dry in air at room temperature for 2 hours.

### *Thermodynamic calculations*

All reaction conditions were, prior to the experiments, evaluated and verified by calculations with the aqueous geochemical simulation program PHREEQC [31] using the Lawrence Livermore National Library database (llnl.dat) and summarized in the SI (SI: Table S2). Missing entries about M-struvites in the database were completed with data from the reported $K_{sp}$ in the literature [32, 33].

### *pH measurements*

The reaction progress was followed in situ by measuring pH as protons were released during precipitation (see eq.[1] and [2]). The pH was continuously measured with a pH electrode connected to a data logger board "DrDAQ" time resolution of 1 s (Pico Technology, Cambridgeshire, UK). The measured pH ranged from the initial 0.1 M DAP solution at around 8.1 to the end of precipitation at 5.6 for the high concentration and 7.2 for the low concentration run. Prior the experiments, the pH electrode was calibrated against buffer solutions.

### *X-ray diffraction*

Powder X-ray diffraction (PXRD, XRD) data from the precipitated powders were collected on a D8 Bruker Diffractometer equipped with a LYNXEYE XE-T detector. Diffraction was measured with Cu Kα radiation (1.5406 Å, 40 kV and 40 mA) from 5-60° using a step size 0.015° (2Θ) and a scanning time of 0.5 s per step.

### *SEM*

The scanning electron microscopy (SEM) characterization was conducted on an FEI XL 30 tungsten cathode scanning electron microscope operated at 20 keV and using a secondary electron detector. Prior to the analysis all samples were coated with a 30 nm thick layer of gold.

### *Dry-TEM*

Dry (i.e. particles directly exposed to high vacuum) transmission electron microscopy (TEM) images were obtained on a Talos F200S Microscope (Thermo Fisher Scientific) operating at 200 kV. A Ceta 16M camera (TEM mode) and a HAADF (high-angle annular dark field) detector (STEM mode; Scanning Transmission electron microscopy) were used to capture the images. Additionally, to collect detailed elemental information, energy dispersive X-ray spectroscopy (EDS) was performed by two side-entry retractable silicon drift detectors (SDD). A counting time of 60 s was applied for point measurements while for the elemental maps 40-50 frames were taken resulting in a counting time of 1 h. The specimen was prepared by mixing the 100 μl of reactant solutions in an Eppendorf tube (1.5 ml) for 5 seconds of reaction time and dropping 10 μl of the sample solution onto a 3 mm holey carbon-coated Cu-grid (Lacey Carbon, 400 mesh). Then, the specimens were left to dry at room temperature for 5 minutes.

### Cryo-TEM

Under "dry" conditions, due to beam damage and high-vacuum, precipitated phases may develop mesopores similar to what is observed in crystalline struvite after thermal treatment [16]. Cryo-TEM allowed us to prevent beam damage artefacts of the material. Cryo-EM imaging was performed with a JEOL JEM-2100 transmission electron microscope (JEOL GmbH, Eching, Germany). Specimens were prepared by casting a 4 μl droplet of sample solution onto lacey carbon-coated copper TEM grids (200 mesh, Electron Microscopy Sciences, Hatfield, PA) and plunge-freezing them into liquid ethane using an FEI vitrobot Mark IV set at 4°C and 95% humidity. Vitrified grids were either transferred directly to the microscope's cryo-transfer holder (Gatan 914, Gatan, Munich, Germany) or stored in liquid nitrogen. All grids were glow-discharged before use. Imaging was carried out at temperatures around -180°C. The TEM was operated at an acceleration voltage of 200 kV, and an objective lens defocus of about 1.5–2 μm was used to increase the contrast. Micrographs were recorded with a bottom-mounted 4·4k CMOS camera (TemCam-F416, TVIPS, Gauting, Germany). The total electron dose in each micrograph was kept below 20 $e^-/Å^2$.

### X-ray Absorption Spectroscopy (XAS)

To evaluate and compare the metal coordination environments in different phases obtained through synthesis, XANES (near-edge X-ray absorption fine structure) and EXAFS (extended X-ray absorption fine-structure) spectroscopy measurements were performed at the BAMline (BESSY-II, Helmholtz Centre Berlin for Materials and Energy Berlin, Germany) [34].

The beam was monochromatized using a silicon double-crystal monochromator (DCM) with a crystallographic orientation of [111]. The size of the beam was 3 mm (l) x 1 mm (h). The measurements were performed at Co-K edge (7709 eV) and Ni-K edge (8333 eV) in transmission geometry, with two ionization chambers as detectors. The excitation energy was varied from 7606 eV to 8678 eV for Co and 8230 eV to 9302 eV for Ni, with varying energy steps. For the pre-edge region, the energy was varied in 10 eV steps; for the region around the edge, energy was tuned first in 0.5 eV steps, then in 1 eV steps and in the EXAFS region with a constant step in the k-space of 0.04 Å$^{-1}$. The associated uncertainties were experimentally determined by measuring the cobalt and nickel metal foils, 10 times each. A value of ±0.3 eV was obtained for both systems. For the measurement, the samples were mixed with boron nitride, placed in polycarbonate hole plates with a thickness of 1 mm and sealed with a polyimide tape (Kapton) on both sides. Before collecting the sample spectra, a cobalt or nickel foil was used as a reference for the respective K edges. The relative energies of the spectra were calibrated to the first inflection point from the first derivative of the cobalt/nickel metal absorption edge.

The resulting EXAFS data were processed by using ATHENA and ARTEMIS. Both programs belong to the main package IFEFFIT (v. 1.2.11) [35]. All crystallographic structures were visualized with VESTA (v. 3.5) [36].

**Results**

The synthesis of transition metal struvites was optimized using different reaction conditions (SI: Tables S1 and S2). The concentrations of the reactant solutions ($Mg^{2+}$, $Ni^{2+}$ and $Co^{2+}$) and initial pH value by adding prior mixing $NaOH_{(aq)}$ or $H_2SO_{4(aq)}$ were varied to explore the phase stability field. The considered reaction conditions were primarily selected to optimize the crystal growth. However, although they do not fully represent the overall complexity of waste streams, the concentration ranges match those known from various real-life scenarios (for instance battery recycling facilities) (Supplementary Note 1 and Table S3).

### *Phase composition in the transition metal phosphate systems*

We explored the potential phase diagrams in the transition metal/DAP systems of M-struvites, M = Mg, Ni and Co (Figure 1). The data points are derived from the XRD measurements (SI: Figure S1 and S2). At all concentration of $Mg^{2+}$, $Ni^{2+}$ and for all the different M/P ratios, pure single-phase Mg or Ni-struvites ($NH_4MgPO_4 \bullet 6H_2O$, $NH_4NiPO_4 \bullet 6H_2O$) are present (Figure 1A-C) without any indicator of amorphous phases. In the Co system at high DAP concentrations of 0.1 M only at a low M/P < 0.4, Co-struvite $NH_4CoPO_4 \bullet 6H_2O$ (COS) precipitated as a pure single phase (Figure 1D). At higher M/P > 0.5, cobalt(II)phosphate octahydrate $Co_3(PO_4)_2 \cdot 8H_2O$ (CPO) is formed as the only phase. In the intermediate interval 0.5 > M/P > 0.4 a binary mixture of CPO and COS is formed (SI: Figure S2A). At lower concentrations of DAP of 0.02 M the transition M/P ratio of Co-struvite to Co-octahydrate is slightly reduced toward lower values < 0.4.

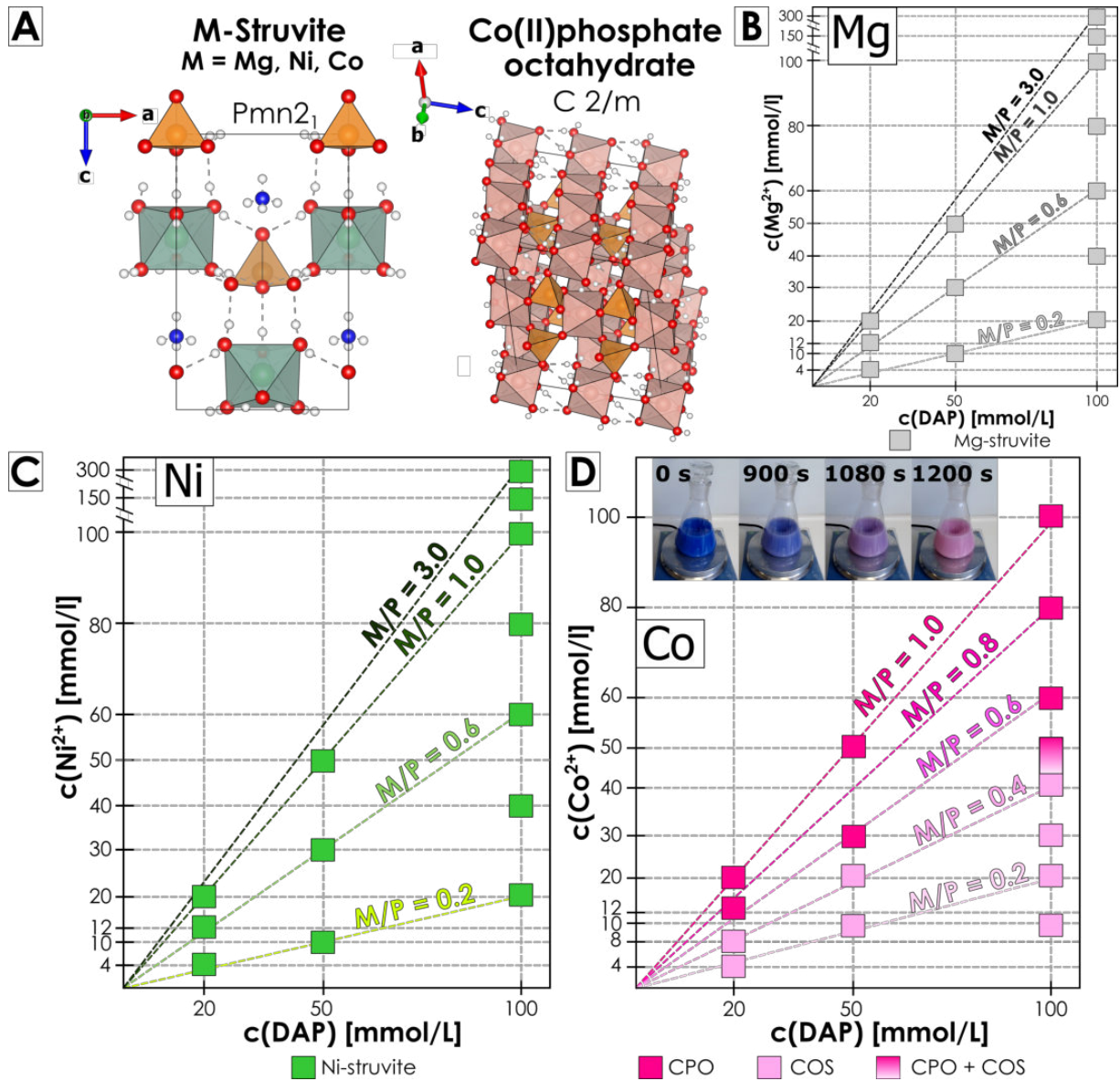

Figure 1: Phase composition diagrams for Mg, Ni, and Co phosphates, which are derived from XRD data (SI: Figures S2-S3); (A) The distinct crystallographic structure of M-struvite $NH_4MPO_4 \cdot 6H_2O$ (010) plane, M = Mg, Ni, Co and Co(II)phosphate octahydrate $Co_3(PO_4)_2 \cdot 8H_2O$ is displayed. For (B) Mg and (C) Ni systems, only struvite was observed. For (D) Co a color change of the aqueous solution from blue to purple to light pink at various times can be observed (see inset); The following symbols and abbreviations are used: CPO = $Co_3(PO_4)_2 \cdot 8H_2O$ cobalt(II)phosphate octahydrate, pakhomovskyite; COS = $NH_4CoPO_4 \cdot 6H_2O$ Co-struvite. Crystal structure colors: white = hydrogen H, blue = nitrogen N, red = oxygen O; orange = phosphorous P; turquoise = metal M = Mg, Ni, Co; pink = cobalt Co; dotted lines hydrogen bonds.

To evaluate how the initial pH impacts the phase stability of M-struvites, different amounts 1-10 ml of 1 M $NaOH_{(aq)}$ and 0.1 M $H_2SO_{4(aq)}$ were added prior to mixing. For all the samples in a pH-series, $c(M^{2+})$ = 0.02 M and c(DAP) = 0.1 M were used to ensure, that the crystals formed within the phase stability field of M-struvite, common for all the systems (M = $Mg^{2+}$, $Ni^{2+}$, $Co^{2+}$). Indeed, for all initial pH adjusted samples with $NaOH_{(aq)}$/$H_2SO_{4(aq)}$ and independent of the metal used, M-struvites formed as pure single phase (SI: Figure S3A-F). We calculated the final equilibrium state of the solution based on the calculations with PHREEQC (SI: Table S2).

## Evolution of precipitation reaction and the role of transitional colloidal phases

In both Ni and Mg systems the respective M-struvite formed as a pure single phase regardless of the reactant concentrations (Figure 1). However, in the case of Co at concentrations of $c(Co^{2+}) > 0.4$ M with $c(DAP) = 0.1$ M other phases precipitated. The phase diagrams (Figure 1) reflect only the final close-to-equilibrium outcome of the crystallization experiments. However, it must be further appended by the observations we made during our studies, which hint about the pathways by which the phases formed. The Co-struvite samples spontaneously decompose in air to Co-dittmarite $NH_4CoPO_4 \cdot H_2O$ (COD), as it was evidenced by the XRD (ambient lab conditions, 1 to 3 days depending on a concentration 0.004 M (t = 1 d) to 0.4 M (t = 3 d)) and was indicated by a color change from light purple to light pink. After several minutes the cobalt phosphate phases changed color from blue, to purple, to light pink. The color change is associated with a phase- and coordination transformation from amorphous to crystalline and octahedron to tetrahedron as discussed later (Figure 1) [37, 38]. The time after which the transformation occurred was highly dependent on the stirring rate, temperature, pH, and concentration of the reactants [39]. In the Ni-struvite system, we also observed a very short-lived apparent colloidal precursor phase, as was hinted by a smooth color transition from light green to emerald. Mg-struvite just showed an initial transparent solution which turned immediately after several seconds turbid due to white precipitates. These indicated complex crystallization pathways of M-struvite should be studied more from point of their evolution rather than their final products.

     Therefore, to establish the sequence of distinct stages during the evolution of potential struvites, we followed the crystallization mechanisms by conducting time-resolved pH measurements. The evolution of pH constituted a proxy for reaction progress, following eqs. [1] and [2]. Consequently, we measured pH vs. time for different metal concentrations $M^{2+}$ for up to 3600 s (SI: Figure S4, Supplementary Note 2). Based on those simple in-situ pH measurements, we found that the transitional metal systems differ significantly from the Mg system for all the concentration runs. The as-observed pH trends and the accompanying color changes of the solution are likely to indicate the presence of long- to short-lived colloidal phases in the transition metal systems on the way to final crystal (Figure S4, Supplementary Note 2). Therefore, as these transitional phases[27, 38, 40-42] affect both the kinetics of the precipitation reaction as well as crystallinity of the final products, it was necessary to investigate them at least at the early stages of nucleation. For this purpose, ex situ dry TEM and cryo-TEM measurements combined with EDS mappings and SAED analysis were performed. Each sample was prepared for a mixing time of 5 seconds (Figure 1), as was established based on the pH measurements. The Mg sample showed at first amorphous round nanoparticles, which rapidly condensed within seconds to μm-sized crystals of needle-shaped Mg-struvite with well-developed faces (Figure 1A and SI: Figure S5). The Ni- and Co-samples exhibited similar amorphous (see SAEDs in Figure 2) round nanoparticles with sizes of ~50 nm and chemical composition very close to M-struvite, which agglomerated to bigger units (Figure 1B-D, SI: Figures S6 and S7). The Ni-nanoparticles and their aggregates appeared more condensed and regular compared to those of Co. Based on the pH trends (SI: Figure S4) it was apparent that the Ni system contained short-lived nanophases present for

several seconds, while for the Co system the nanophases were long-lived and stable up to hours.

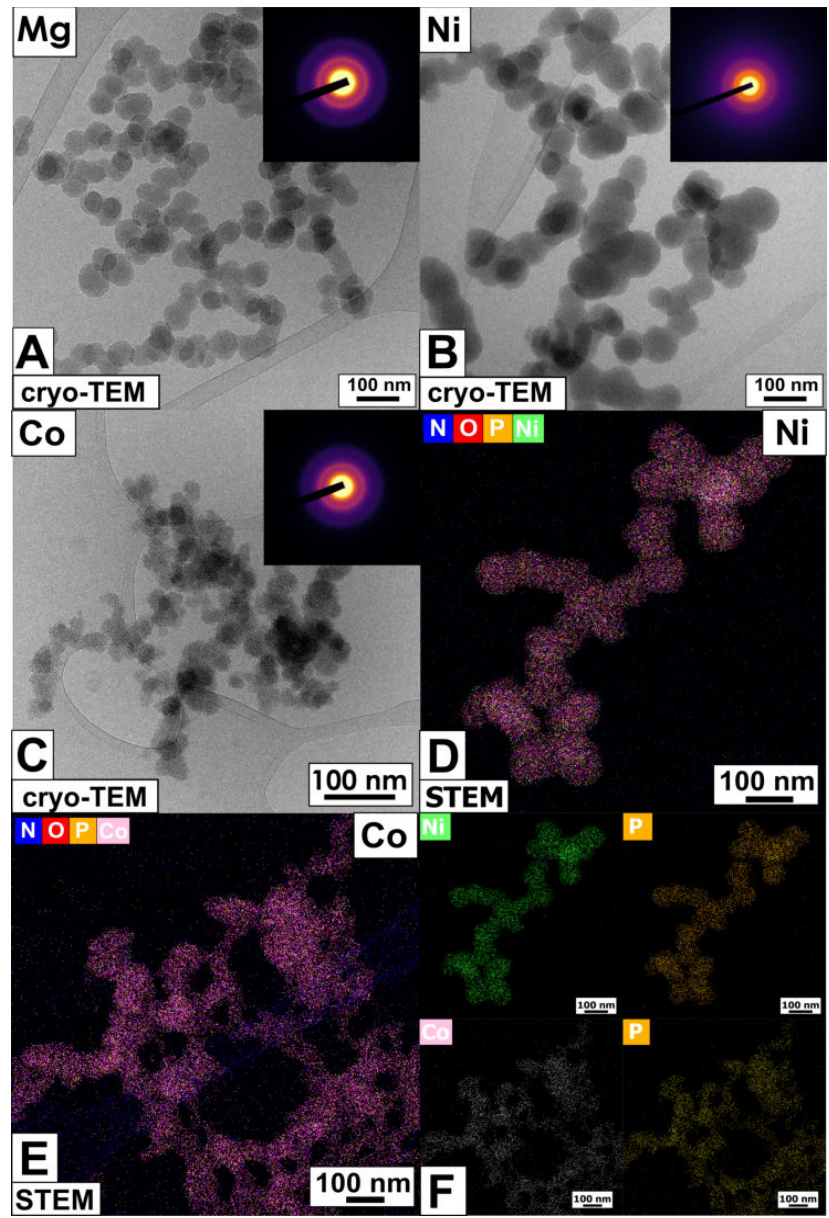

Figure 2: cryo-TEM images of (A) Mg sample after 5 seconds mixing with an inset of the SAED pattern; (B) cryo-TEM images of Ni sample after 5 seconds mixing with an inset of the SAED pattern; (C) cryo-TEM images of Co sample after 5 seconds mixing with an inset of the SAED pattern; All SAED patterns clearly show the amorphous nature of the observed particles; (D) EDS mapping of amorphous Ni-PO$_4$-phases occurring after 5 seconds of mixing; (E) EDS mapping of amorphous Co-PO$_4$-phases occurring after 5 seconds of mixing; (F) EDS mappings of selected elements (Ni, Co and P) in the occurring Ni -and Co-PO$_4$ phases after 5 seconds of mixing; Supporting detailed STEM images, additional EDS mappings of more elements and EDS point measurements can be found in the SI (SI: Figures S6-S7);

### *Crystal engineering of transition metal phosphates*

To quantify how the changes in concentration of transition metal $M^{2+}$ and DAP, as well as the initial pH, affect the morphology and size of the crystallites, SEM analysis was performed on crystals synthesized for various conditions (Figure 3, SI: Figure S8-S13). Based on SE images, the lengths of intact crystals were measured. Exemplary size distribution histograms of Mg-, Ni- and Co-struvite can be found in the SI (SI: Figure S14). A strong dependency of the morphology and size of the crystals was observed in the Ni-system in respect to the M/P ratio (Figure 3A, SI: Figure S8 & S9). Ni-struvite morphologies with an M/P ratio ≥ 1 (high $c(Ni^{2+})$ compared to low c(DAP)) showed an anhedral rounded appearance and numerous agglomerated crystal nuclei with a crystallite size of ~20 μm (length from edge-to-edge) at c(Ni)=c(DAP)= 100 mM (SI: Figure S14B). At low M/P ratios ≤ 1 (low $c(Ni^{2+})$ compared to high c(DAP)) the crystals exhibited a more euhedral elongated X- or "coffin-like" shaped habit with crystallite size of 50 μm at c(Ni) = 20 mM and c(DAP) = 100 mM. The X-shaped habit of the crystals occurred at low M/P ratios < 0.5 and high c(DAP) ≥ 50 mM while the "coffin-like" morphology appeared at low M/P ratios < 0.5, but at low c(DAP) ≤ 50 mM. High $Ni^{2+}$ concentrations at a constant DAP concentration favored numerous crystallites of smaller size with an anhedral appearance, while high DAP concentrations at a constant $Ni^{2+}$ concentration led to an increase in the crystallite size and idiomorphism of the crystallites. Adding $NaOH_{(aq)}$ and the coupled increase of pH favored the formation of M-struvite because protons were released through the precipitation reaction which counteracted the pH effect of $OH^-$ ions (eq.[1]) while adding $H_2SO_{4(aq)}$ inhibited the precipitation. Adding $NaOH_{(aq)}$ to the Ni-struvite system alter the crystal morphology strongly and decreased the crystallite size from ~50 μm to ~10 μm (grey inset in Figure 3A, SI: Figure S10). In addition, it changed the crystal habit to an orthorhombic prism appearance. Adding $H_2SO_{4(aq)}$ affected the surface of the crystals strongly, although under these conditions the crystallite size and the X-shaped-like morphology were not influenced in comparison with the pH-unadjusted sample at the same $M^{2+}$ and DAP concentrations.

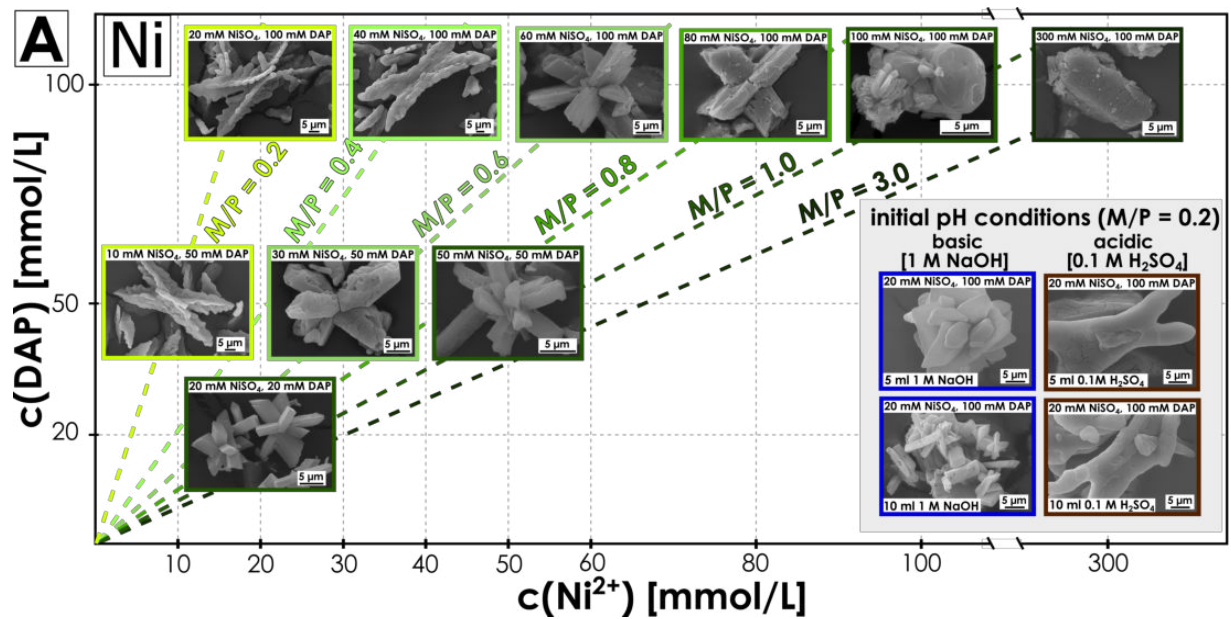

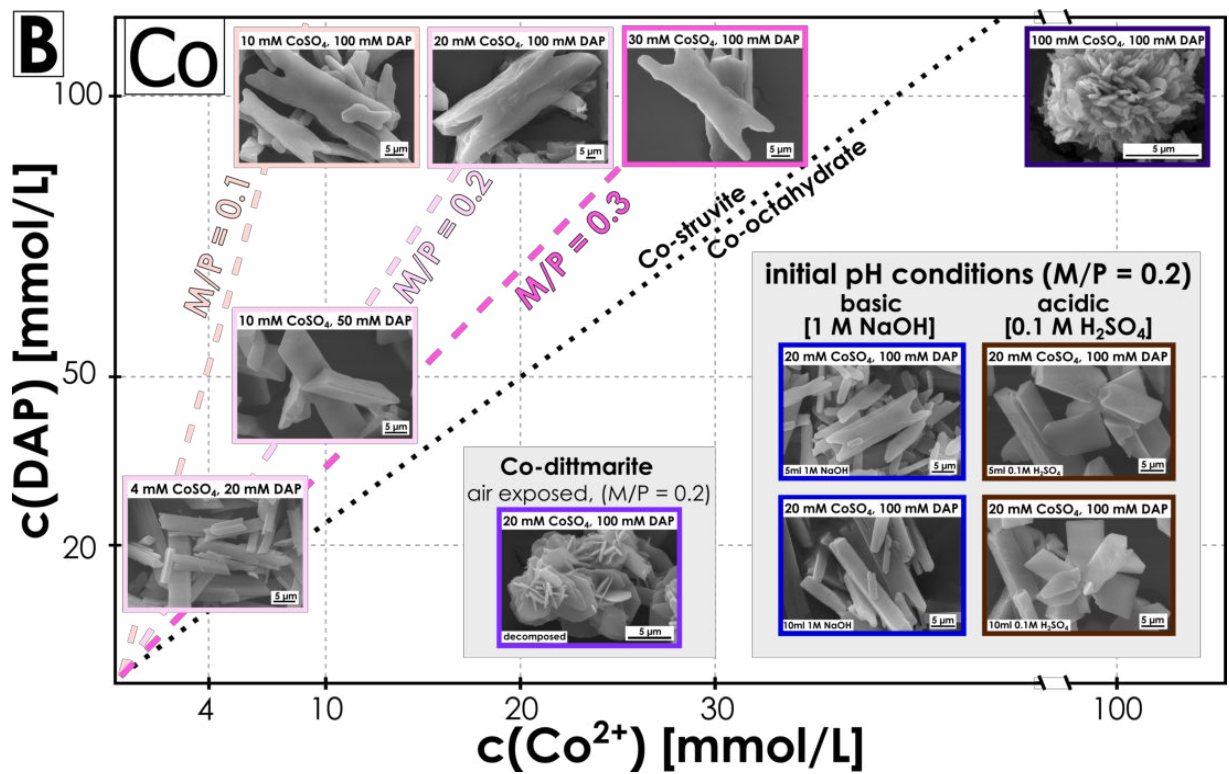

Figure 3: c($M^{2+}$) vs. c(DAP) plot with (A) green (for Ni) and (B) pink (for Co); color frames and dotted lines indicate the distinct M/P ratio of the sample; lower right left corner: pH samples with a M/P ratio = 0.2 (0.02 M NiSO$_4$/CoSO$_4$, 0.1 M DAP) with different initial amounts of 1 M NaOH$_{(aq)}$ (blue frames) or 0.1 M H$_2$SO$_{4(aq)}$ (brown frames); in air Co-struvite decomposed to Co-dittmarite with a M/P ratio = 0.2. and is displayed in (B).

Due to the limited stability of Co-struvite, only samples with the M/P ratio below 0.4 could be analyzed. Above the ratio of 0.4 only Co-octahydrate was stable. At a constant ratio the crystallite size increased from ~15 μm for a 4 mM Co$^{2+}$ and 20 mM DAP to ~45 μm for a 20 mM Co$^{2+}$ and 100 mM DAP (SI: Figure S14C). The Co system followed a similar, but less-pronounced correlation to the one seen for the Ni. High DAP concentrations favored the formation of large euhedral crystals while higher Co$^{2+}$ contents led to small anhedral crystals. Similarly, to Ni-struvite, Co-struvite exhibited X-shaped crystal morphologies at high c(DAP) ≥ 50 mM and low M/P ratios < 0.3, but the crystallographic planes were anhedrally developed (Figure 3B, SI: Figure S11). The "coffin-like" habit occurred at low M/P ratios < 0.3 and low c(DAP) ≤ 50 mM. Co-octahydrate showed a rough two-dimensional flake- or sheet-shaped morphology of the crystals and formed big crystal agglomerates of several μm. Co-dittmarite demonstrated a similar habit of regularly arranged plates, also reported in other studies [37]. By increasing the pH with NaOH the crystal habit changed from a X-shaped morphology to elongated rod or prism-like habit (grey insets in Figure 3, SI: Figure S12). Here, the 5 ml 1 M NaOH$_{(aq)}$ sample indicates remarkably the transition between those two morphologies. 10 ml NaOH$_{(aq)}$ decreased the crystallite size from ~60 μm to ~20 μm. Adding H$_2$SO$_{4(aq)}$ favored the formation of tabular-shaped Co-struvites with slightly decreased size of 40 μm. All trends in terms of crystal morphology and size for Ni and Co are summarized in Figure 4.

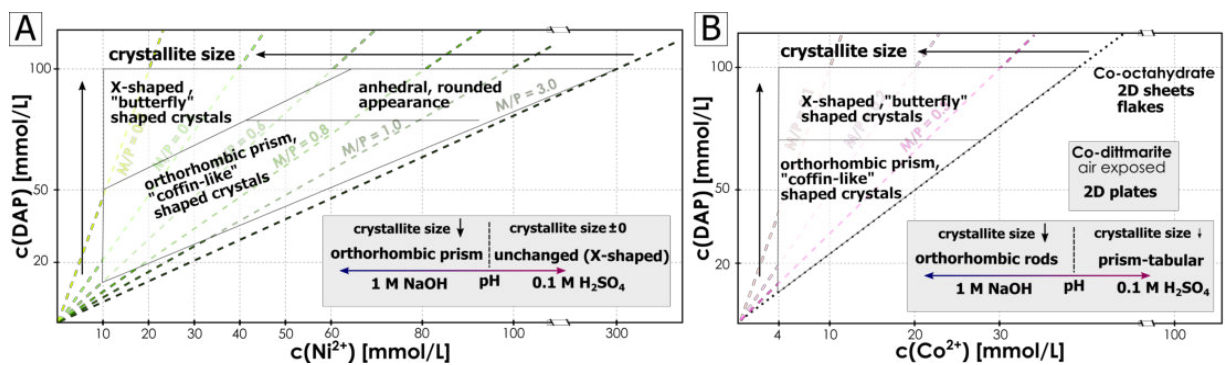

Figure 4: Crystallization trends in the c($M^{2+}$) vs. c(DAP) plots with (A) green (for Ni) and (B) pink (for Co) dotted lines indicating the distinct M/P ratio of the sample. Crystal morphology fields are displayed in the triangles. The crystallite size increases with higher DAP and lower $M^{2+}$ contents (see direction of the arrows); Adding $NaOH_{(aq)}$ decreased the crystallite size and favored the formation of orthorhombic prisms. Co-octahydrate exhibits a 2D sheet or flake morphology. Co-dittmarite (formed due to air exposure) shows a 2D plate habit. In the Ni-system the crystallite size remained unchanged and while in the Co-system the crystals decreased slightly in size and habit to more tabular-shaped crystals.

Mg-struvite is a relatively extensively investigated mineral phase [16, 40, 43], and therefore, it was synthesized to serve as a reference for transition metal phosphates (SI: Figure S13). At high concentrations of DAP and $Mg^{2+}$ = 0.1 M, struvite formed an X-shaped crystal habit with a crystallite size of ~80 µm, while at low concentrations orthorhombic prisms were observed and the crystallite size decreased slightly to ~50 µm (SI: Figure S14A). In the case when we adjusted conditions to be initially acidic (adding 10 ml of 0.1 M $H_2SO_4$ with a pH = 1.00) the struvite crystals were comparable in terms of shape and size to those observed in the pH unadjusted samples. However, the surface exhibited features of acidic leaching. Under basic conditions adding 10 ml of 1 M $NaOH_{(aq)}$, the size of the crystallites got reduced to 5 µm and the shape changed to more rounded orthorhombic prisms, but in contrast to acidic conditions the crystal surface were visually not affected. These results demonstrate a controllable and tunable adjustment of the crystallite size, morphology, and composition of M-struvites through changes in the reaction conditions.

*Stability of phases from the point of coordination environment*

Crystalline M-struvite form in both transitional metal systems ($Ni^{2+}$, $Co^{2+}$), but they exhibit different stability. Namely, Co-struvite decomposes in air to Co-dittmarite while Ni-struvite remains stable. The observed instability of crystalline Co-struvite is quite surprising, and as we show below, is not caused e.g. by oxidation of $Co^{2+}$. Instead, these differences can be explained in the context of the local coordination environment of the $MO_6$ octahedron in the struvite structure. To evaluate how the stability of Ni and Co phosphates is related to their coordination environment in the crystal structure and how they differ between each other, EXAFS measurements were performed (Figure 5, SI: Figure S15).

In general, XAS analysis reveal quantitative changes in oxidation state, in the local coordination geometry/symmetry and bond distances among the different Co- and Ni-phases. Especially the intensity of the pre-peak (XANES) which originates from the *1s-3d* transitions expresses the degree of centrosymmetry and coordination environment in the vicinity of the considered ions (SI: Figure S15, Table S4). As only *3d-4p* hybridization enables parity allowed transitions the extent of those is strongly correlated with the coordination geometry. Near-ideal centrosymmetric coordination shows very low pre-peak intensities while more distorted geometries exhibit higher intensities. Quantitative real-space fittings on the radial distribution function reveal the bond distances in the first coordination environment around the $Co^{2+}$ and $Ni^{2+}$ metal cation and can be linked to the crystallographic structure (Figure 5).

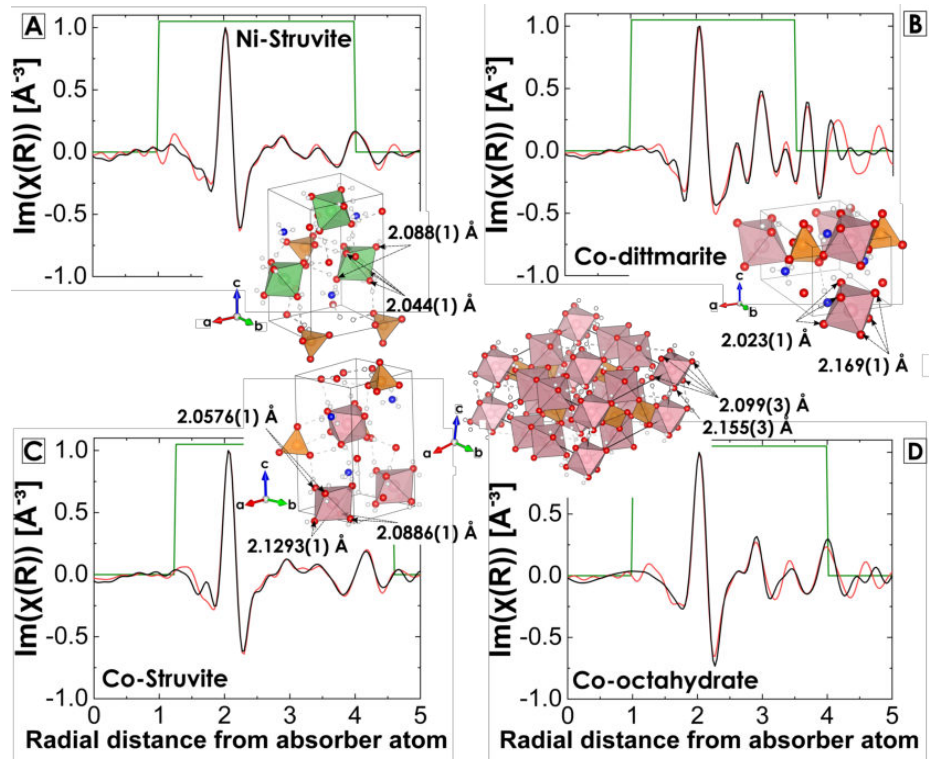

Figure 5: Ni- and Co K-edge EXAFS data shown in real space for (A) $NH_4NiPO_4·6H_2O$ space group: $Pmn2_1$, R = 0.005; (B) for $NH_4CoPO_4·H_2O$ space group: $Pmn2_1$, R = 0.008; (C) $NH_4CoPO_4·6H_2O$ space group: $Pmn2_1$, R = 0.012; and (D) $Co_3(PO_4)_2·8H_2O$ space group: $C2/m$, R = 0.005; experimental data: red, fit results: black; fit window: green.

The XAS spectra (SI: Figure S15) clearly prove that Co and Ni are exclusively present as the bivalent species in all samples, as is evidenced by comparing with multiple $M^{2+}$ standards (only $MSO_4·nH_2O$ is shown for clarity). No indication for incorporated of oxidized $Ni^{3+}$ or $Co^{3+}$ was observed in the powders. Quantitative pre-peak integration results are summarized in Table S4. As it is indicated by a minimal pre-peak area ($A_{pp}$= 0.0298(3)) and a maximized intensity in the white line, Co-struvite as well as Ni-struvite ($A_{pp}$= 0.0281(2)) have a more centrosymmetric octahedral coordination environment compared to Co-dittmarite and Co-octahydrate ($A_{pp}$=0.0363(4) and $A_{pp}$=0.0355(4)) (SI: Table S4). After FT of $\chi(k)$ into R-space radial distances from the absorber atom can be observed (SI: Figure S15C). At around R ≈ 1.9 Å all samples show a maximum which is related to the degeneracy of single scattering path of the Ni-O or Co-O bond length. In the case of single scattering, we have a direct measure of the coordination number (CN). The peak at R ≈2.7 Å in CPO and COD fits to the Co-P single scattering path, which is absent in the M-struvites. In CPO and COD the $CoO_6$ octahedrons are linked to four (CPO) or five (COD) $O^{2-}$ of $PO_4$ tetrahedrons (short distance of Co-P) while in the struvite structure they are isolated from the phosphate tetrahedron through coordination of six $O^{2-}$ of water molecules (long distance of Ni-P and Co-P, SI: Figure S16).

The slight color change among COS, COD and CPO of the chemical compounds of Co is related to the number and the distortion of the coordinating water molecules. The peak at R ≈ 3.8 Å consists of several single and multiple scattering paths such as Co-N, Co-H, Co-P, Co-O-O. All fitting parameters are shown in SI: Tables S5-S8. By calculating the mean bond distance between $M^{2+}$ and $O^{2-}$ in the first coordination sphere and the occupying sphere volume, a quantitative comparison of the coordination sphere is possible. The M-struvites (M-struvite Ni: 2.059(6) Å Co: 2.091(1) Å) demonstrate slightly lower mean bond distances than COD and CPO (CoD: 2.096(1) Å, CPO: 2.11(2) Å). The fitted bond distances are visualized with the calculated fits in Figure 5 for the different compounds (SI: Tables S5-S8).

**Discussion**

*General phase selection, stability and transformations*

In broad terms, whether an actual transition metal struvite might be a probable phase is determined by the M cation's charge and radius compared to $Mg^{2+}$. Considering the ionic radii of $Co^{2+[VI]}$(0.75 Å, high state) and $Ni^{2+[VI]}$ (0.69 Å) compared to $Mg^{2+[VI]}$ (0.72 Å), all ions have similar radii so they could be potentially substituted one for another in the struvite structure $Pmn2_1$ (Figure 1A) . However, in practice it is apparent that the crystallization pathways play a crucial role in the phase selection. In all the considered chemical systems crystalline M-struvites could be obtained, but the conditions and the nucleation pathways toward the final crystal varied among different cations. For all three metal cations, single-phase M-struvite was obtained within a defined reaction window (Figure 1B-D), while specifically for Co at higher M/P ratios, phosphate octahydrate (CPO) was crystallizing instead of struvite. It was evident that for both Ni and, even more clearly for Co, the crystalline phases formed through transformation of amorphous phases accompanied by a color change. This transition of colors is highly significant because it points to specific meta-stable reaction products. For all systems amorphous phases play a role in the nucleation of struvite, as we discuss below. More significantly, in the case of Co, transitional amorphous phases form also when CPO is the more stable phase. During the formation of CPO at high reactant concentrations, since ammonium is the most abundant species in the aqueous solution, the dissociation equilibrium of $NH_3/NH_4^+$ with the release of protons dominates the pH. The remaining less abundant phosphate due to the M/P ratio in CPO of 3:2, and the coupled dissociation to $H_xPO_4^{3-x}$ species, with the removal of protons can only compensate to a limited extent. The jump of the pH after 1800 s (SI: Figure S4) could be influenced by the formation of $Co(NH_3)_xH_2O_{6-x}^{2+}$ aqua complexes indicated by a light pink color [44, 45]. The 0.1 M concentrated Ni and Mg samples do not exhibit any late reduction of pH as most of the ammonium is bound in the struvites, and probably no amorphous metal phosphate phase is stable over a significantly longer period (> several mins), although it is briefly yet clearly present for Ni.

*Significance of amorphous phases for nucleation of struvite*

Interestingly, Mg-struvite and Ni-struvite seem to have higher stability compared to Co-struvite, because no other phases precipitate in solution. Looking at the control of reaction time in the dry and cryo-TEM samples, it must be accounted that probably the chemical compounds may have reacted longer than their mixing time in the solution would indicate. Since the droplets of the mixed solution had to be quenched by blotting on the TEM grid, the sample progressed at least a few seconds further in the reaction. In contrast, in cryo-mode the pipetted solution got near-immediately frozen after the injection into the vitrobot, effectively quenching chemical reactions. Therefore, we suggest that the real reaction time $t_{reaction\ time}$ of the reactants is closer to the mixing time $t_{mix}$ in the cryo-samples due to a close-to-native sample preservation, whereas in the dry TEM samples a few more seconds passed until the droplet dried completely $t_{reaction\ time} \leq t_{mix,cryo} < t_{mix,dry}$. However, from pH trends (SI: Figure S4), dry and cryo-TEM measurements combined with SAED, Mg- and Ni-struvite show short-lived amorphous nanophases (Figure 1, SI: Figure S5 and S6), which form in the early stages of precipitation within seconds, or faster (in Mg-struvite). These regular amorphous agglomerates of spherical particles aggregate and transform at higher reaction times to a precursor "unit" of a crystal (Figure 1B, SI: Figure S6). The Co-system exhibits metastable long-lived (minutes to hours) amorphous phases on the way to the final crystalline Co-struvite, which is correlated with a color change from purple to light pink. Here, the $Co^{2+}$ changes its configuration from tetrahedral to octahedral coordination which leads to a slightly different color absorption [37, 38]. Similarly to the Mg- and Ni-samples, Co formed round amorphous nanoparticles (Figure 1C, SI: Figure S7). The condensation and densification to a proto-crystalline unit took place throughout a longer time period (minutes to hours) than in the Mg- or Ni-system. It may explain why only a weak correlation between concentration of $Co^{2+}$/DAP and the resulting crystal habit is observed compared to Ni and Mg. Based on the results, it is clear that the crystallization of Mg- and transition metal struvite follows a non-classical nucleation pathway [46, 47] which involves transitional amorphous precursor phases on the way to the final crystalline product (Figure 6). The amorphous colloidal nanoparticles form within seconds, aggregate and transform within minutes of reaction time. These aggregates transform to a proto-crystalline nucleus which facilitate the next crystallisation of new nuclei. The life time of the transitional amorphous nanophases in M-struvite is influenced by the involved metal cation $M^{2+}$ as $Mg^{2+}$, $Ni^{2+}$ seem to react faster to a final crystal than $Co^{2+}$. In the final step depending on the reaction conditions these crystal nuclei develop into different morphologies and sizes of crystals.

*Crystal morphology and phase*

Following the formation of the crystals, a higher supersaturation leads to a higher nucleation rate which provides numerous crystal nuclei of smaller crystallite sizes (saturation index > 1, $K_{sp}$ of M-struvite $K_{sp}^{Mg} = 10^{-13.36} > K_{sp}^{Ni,Co}$)[32, 33]. At low supersaturations only a few crystal nuclei grow significantly due to a lower nucleation rate. Based on observations, the $Ni^{2+}$ and $Co^{2+}$ concentration controls the struvite morphology in terms of the supersaturation and reaction kinetics. In addition, by increasing initial metal concentration, the presence of metal aqua complexes becomes more significant, where the complexes influence the crystallization through positive feedback for instance by consuming available ammonium ions. Higher concentrations of DAP correlate with the apparent elongation and increasing crystallite size of the crystals at a constant M/P ratio (increased supersaturations). Similar effects were observed for the crystallization of Mg-struvites [43, 48]. Since in DAP two moles of $NH_4^+$ compared to one mole of $PO_4^{3-}$ are present, the effect of higher DAP concentrations on the crystals is mostly related to higher concentrations of ammonium. In Mg-struvite higher concentrations of ammonia promoted the selective growth in direction of [00i] with i = 1-4 along the c-axis due to a higher density of these electropositive groups ($NH_4^+$) on this facette. $PO_4^{3-}$ and $M(H_2O)_6^{2+}$ favor the crystal growth of the (00i) facettes with i = $\overline{1} - \overline{4}$. Based on our results, transition metal struvite indicates a similar trend in its crystal growth. Looking at the morphology, round crystals have the lowest A/V (surface area/volume) ratio compared to rods and X-shaped ones. In the M/P diagram supersaturations would occur as n/x curves n > 0. Higher supersaturations lead to more rounded shaped, while lower supersaturations lead to X-or rod- shaped morphologies.

In all systems an initial basic pH by adding $NaOH_{(aq)}$ decreased the crystallite size significantly and stabilized a more rod/prism-like morphology of the crystals. An initial acidic pH by adding $H_2SO_{4(aq)}$ remained the crystallite size practically unchanged in the Ni-system, but the crystal surface was clearly affected by the acid. In the Co-system the crystallite size decreased slightly to ~10 μm and the morphology changed to more tabular shapes. A slight basic initial pH (~8-9) seems to be the optimal value for maximized growth and euhedral habit in both systems. The Co-samples treated with sulfuric acid exhibited a short presence of the transitional amorphous colloidal phases of 10-15 mins as well as they decomposed faster to Co-dittmarite within several hours compared to the pH unaffected samples. Since only 0.1 M $H_2SO_{4(aq)}$ was added initially in the amount sufficient not to dissolve the crystals completely, the absolute change in pH, |ΔpH|, was lower and asymmetric than the one induced by adding higher concentrated 1 M $NaOH_{(aq)}$ to the system. As an increase in the pH (a more basic pH) shifts the equilibrium to the side of the products in eq.[1], a high number of crystal nuclei are formed and grow simultaneously. Consequently, as many crystals of a low size have a higher surface area compared to few crystals of a larger size, these nuclei would grow only to a limited extent. Therefore, these crystals are much smaller than in the pH unadjusted runs where only few crystals grew large. Low-concentration sulfuric acid already reduces the lifespan of the transitional colloidal nanophases and the final crystalline Co-struvite significantly. From the point of view of saturation index (SI) and ion activity product (IAP), an initial basic pH increases the $SI_{M-struvite}$ as the activity of $[PO_4^{3-}{}_{(aq)}]$ increases. However, this shift of the equilibrium to the product side is limited to a certain pH, because at a high pH (for Mg-struvite pH = 11) the presence of $NH_{3(aq)}$ and $M(OH)^-{}_{(aq)}$ species reduces the activities of $NH_4^+{}_{(aq)}$ and $M^{2+}$ and therefore compensates for the increase of activity of $PO_4^{3-}{}_{(aq)}$ [43].

*Phase stability from a point of view of a metal coordination environment*

Co-struvite exhibits low stability compared to Ni-struvite as it spontaneously decomposes in air to Co-dittmarite or is replaced at higher concentrations of $c(Co^{2+})$ > 0.4 M by CPO as the most stable phase. This is quite surprising, in particular in the case of fully developed and supposedly thermodynamically stable crystals. Hints on these stability properties are hidden in the metal coordination environment. The $CoO_6$ octahedron in the struvite structure shows a higher degree of distortion compared to $NiO_6$ in Ni-struvite, but a lower one compared to the other Co-phases (pre-peak area fits in SI: Table S4, Figure S15A, 15B). The volume and the distortion of the $CoO_6$ octahedron could be "straightened" to allow for stability of the M-struvite structure. But, the Co-struvite tends to decompose to Co-dittmarite. Here, the Co-dittmarite structure allows for a more distorted coordination environment of the central metal cation since it is coordinated by five corner-linking $O^{2-}$ anions of the phosphate tetrahedrons and one $O^{2-}$ anion from the remaining crystal water (SI: Figure S16). This assumption takes into account the high stability of Co(II) phosphate octahydrate at high M/P ratios > 0.4 (Figure 3B, SI: Figure S1), where $Co^{2+}$ is also bound in a similar distorted octahedral coordination as in Co-dittmarite (SI: Figure S15, Table S4). As $H_2O$ behaves as a weak-field ligand for most of the bivalent transition metal ions, the high spin configuration is realized in their complexes. Looking at the Irving-Williams thermodynamic series of high spin first row transition metals in the bivalent oxidation state the stability of metal complexes decreases in the order $Mn^{2+} < Fe^{2+} < Co^{2+} < Ni^{2+} < Cu^{2+} > Zn^{2+}$ regardless of the type of coordinated ligands [49, 50]. $Co^{2+}$ as a $d^7$ ion shows a strong Jahn-Teller effect in octahedral coordination with weak ligands like $H_2O$ (high-state) compared to $Ni^{2+}$ with a $d^8$ orbitals as more unpaired electrons occur in this electron configuration. In agreement with our results, other studies demonstrated a distorted non-centrosymmetric octahedral coordination of $Co^{2+}$ as $d^7$ ion in Co-doped Ni-struvite while $Ni^{2+}$ as a $d^8$ ion exhibits a near ideal cubic octahedral coordination based on electron spin resonance spectroscopy [51]. Therefore, we suggest that the different stabilities of Ni- and Co-struvite are mainly influenced by the electronic d-configuration and their corresponding ionic radii for the struvite structure $Pmn2_1$.

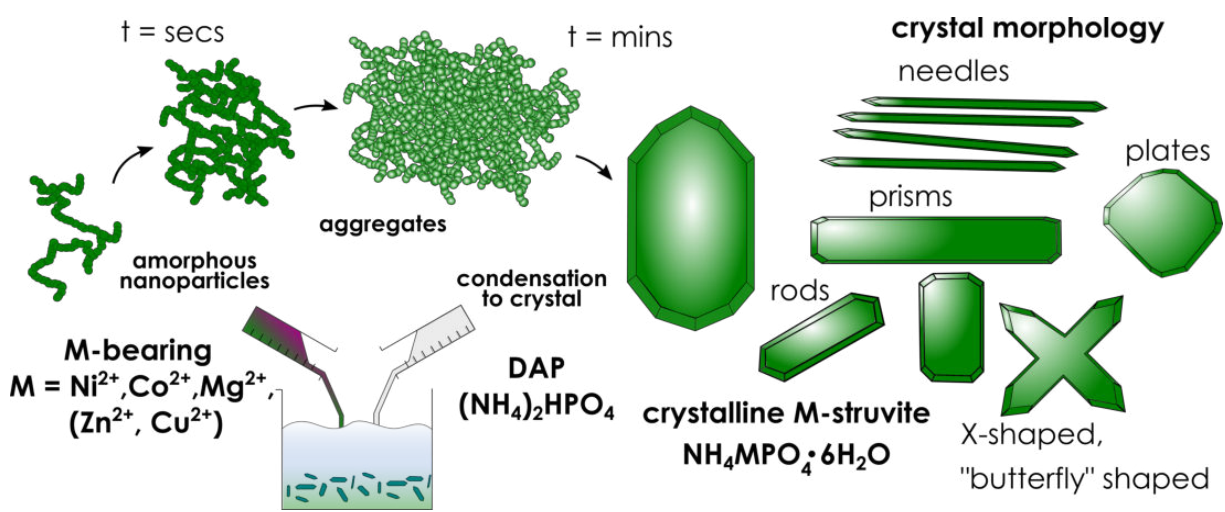

Figure 6: Visualization of the formation of different transition metal struvite crystals through a non-classical crystallization route including amorphous nanoparticles. Stages according to the labels.

**Conclusion**

Our investigation of transition metal struvites as potential systems for multicomponent precipitation led to several crucial insights. Ni-struvite is stable at multiple concentrations of the reactants and at different metal/phosphate (M/P) ratios. Co-struvite precipitates at an M/P ratio of 0.4, whereas above this value only Co(II)-phosphate octahydrate is stable. All M-struvites form amorphous precursor nanophases on the way to the final crystalline product but their kinetics differ majorly. Control of different morphologies and sizes of transition metal struvite can be achieved depending on the crystallization conditions. Transition metal phosphates appear to form through non-classical nucleation pathways, which involve multiple amorphous nanophases during precipitation. Due to the low solubility product and their controlled precipitation through adjusting the reaction conditions transition metal struvite is a promising recovery material for extracting $NH_4^+$, $PO_4^{3-}$ and transition metals at simultaneously, out of synthetical industrial waste waters or sludges. Furthermore, the possibility to specifically set/tune the morphology and properties opens the prospect of up-cycling materials directly as raw materials for applications (e.g. electrocatalysis).

## Acknowledgements

We acknowledge Uwe Reinholz, Martin Radtke and Kirill Yusenko for their support at the BAMline beamline of BESSY II. We acknowledge BAM and Helmholtz-Zentrum Berlin (HZB) for providing us with the beamtime at BESSY II. We thank Carsten Prinz for TEM measurements.

**Graphical abstract**

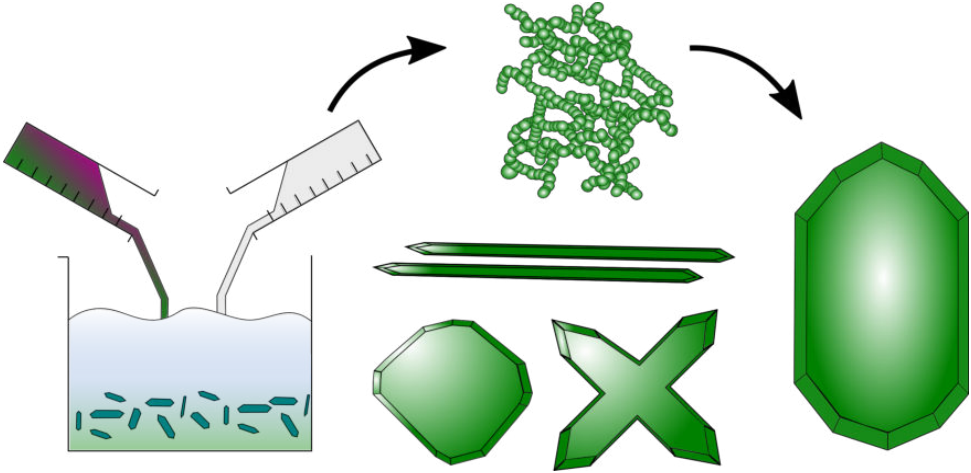

# Supporting Information for:

## Ni- and Co-struvites: Revealing crystallization mechanisms and crystal engineering towards applicational use of transition metal phosphates

Stephanos Karafiludis[1*], Ana Guilherme Buzanich[1], Zdravko Kochovski[2], Ines Feldmann[1], Franziska Emmerling[1,3], and Tomasz M. Stawski[1**]

[1]Federal Institute for Materials Research and Testing, Richard-Willstatter-Straße 11, 12489 Berlin, Germany

[2]Department for Electrochemical Energy Storage, Helmholtz-Zentrum Berlin for Materials and Energy, Hahn-Meitner Platz 1, 14109 Berlin, Germany

[3]Department of Chemistry, Humboldt-Universität zu Berlin, Brook-Taylor-Straße 2, 12489 Berlin

Corresponding authors: **tomasz.stawski@bam.de; *stephanos.karafiludis@bam.de

**List of Tables**

**List of Figures**

**Contents**

Table S1: Detailed reaction conditions for the precipitated samples.

| c(NiSO$_4$) [M] | c(DAP) [M] | M/P ratio | c(CoSO$_4$) [M] | c(DAP) [M] | M/P ratio | c(MgCl$_2$) [M] | c(DAP) [M] | M/P ratio |
|---|---|---|---|---|---|---|---|---|
| 0.004 | 0.02 | 0.2 | 0.004 | 0.02 | 0.2 | 0.004 | 0.02 | 0.2 |
| 0.01 | 0.05 | 0.2 | 0.008 | 0.02 | 0.4 | 0.01 | 0.05 | 0.2 |
| 0.012 | 0.02 | 0.6 | 0.01 | 0.05 | 0.2 | 0.012 | 0.02 | 0.6 |
| 0.02 | 0.02 | 1 | 0.012 | 0.02 | 0.6 | 0.02 | 0.02 | 1 |
| 0.02 | 0.1 | 0.2 | 0.02 | 0.02 | 1 | 0.02 | 0.1 | 0.2 |
| 0.03 | 0.05 | 0.6 | 0.02 | 0.1 | 0.2 | 0.03 | 0.05 | 0.6 |
| 0.04 | 0.1 | 0.4 | 0.03 | 0.05 | 0.6 | 0.04 | 0.1 | 0.4 |
| 0.05 | 0.05 | 1 | 0.04 | 0.1 | 0.4 | 0.05 | 0.05 | 1 |
| 0.06 | 0.1 | 0.6 | 0.042 | 0.1 | 0.42 | 0.06 | 0.1 | 0.6 |
| 0.08 | 0.1 | 0.8 | 0.046 | 0.1 | 0.46 | 0.08 | 0.1 | 0.8 |
| 0.1 | 0.1 | 1 | 0.05 | 0.05 | 1 | 0.1 | 0.1 | 1 |
| 0.15 | 0.1 | 1.5 | 0.05 | 0.1 | 0.5 | 0.15 | 0.1 | 1.5 |
| 0.3 | 0.1 | 3 | 0.054 | 0.1 | 0.54 | 0.3 | 0.1 | 3 |
| | | | 0.058 | 0.1 | 0.58 | | | |
| | | | 0.06 | 0.1 | 0.6 | | | |
| | | | 0.08 | 0.1 | 0.8 | | | |
| | | | 0.1 | 0.1 | 1 | | | |
| | | | 0.15 | 0.1 | 1.5 | | | |
| | | | 0.3 | 0.1 | 3 | | | |

Table S2: PHREEQC calculations of absolute pH values in solution with 100 ml 0.1 M DAP and 100 ml of 0.02 M $M^{2+}$ (M = Mg, Co, Ni) with different amounts of 1 M $NaOH_{(aq)}$(pH = 13.97) and 0.1 M $H_2SO_{4(aq)}$ (pH = 1.00), the reference at the right column without any initial pH adjustment (0.02 M metal-bearing solution and 0.1 M DAP) and the deviation from it in the ΔpH column.

| Experimental conditions | | M/P ratio = 0.2 | | c(MSO₄) [mM] = 20 | c(DAP) [mM] =100 | |
|---|---|---|---|---|---|---|
| Reference without any pH adjustment | amount of 1M $NaOH_{(aq)}$ [ml] | calculated $pH_{eq}$ | ΔpH | amount of 0.1M $H_2SO_{4(aq)}$ [ml] | calculated $pH_{eq}$ | ΔpH |
| **Ni-struvite** | 1 | 7.03 | 0.16 | 1 | 6.82 | -0.05 |
| **$pH_{eq}$ = 6.87** | 2 | 7.21 | 0.33 | 2 | 6.78 | -0.10 |
| | 3 | 7.40 | 0.53 | 3 | 6.73 | -0.15 |
| | 4 | 7.66 | 0.78 | 4 | 6.68 | -0.20 |
| | 5 | 8.04 | 1.17 | 5 | 6.63 | -0.25 |
| | 6 | 8.47 | 1.60 | 6 | 6.57 | -0.30 |
| | 7 | 8.78 | 1.90 | 7 | 6.52 | -0.35 |
| | 8 | 8.99 | 2.12 | 8 | 6.46 | -0.42 |
| | 9 | 9.17 | 2.30 | 9 | 6.41 | -0.47 |
| | 10 | 9.32 | 2.45 | 10 | 6.35 | -0.53 |
| **Co-struvite** | 1 | 7.18 | 0.16 | 1 | 6.98 | -0.04 |
| **$pH_{eq}$ = 7.02** | 2 | 7.37 | 0.35 | 2 | 6.94 | -0.08 |
| | 3 | 7.60 | 0.58 | 3 | 6.90 | -0.12 |
| | 4 | 7.89 | 0.87 | 4 | 6.86 | -0.16 |
| | 5 | 8.19 | 1.17 | 5 | 6.83 | -0.19 |
| | 6 | 8.38 | 1.36 | 6 | 6.79 | -0.23 |
| | 7 | 8.51 | 1.49 | 7 | 6.76 | -0.26 |
| | 8 | 8.59 | 1.57 | 8 | 6.72 | -0.30 |
| | 9 | 8.66 | 1.64 | 9 | 6.67 | -0.35 |
| | 10 | 8.75 | 1.73 | 10 | 6.61 | -0.41 |
| **Mg-struvite** | 5 | 8.86 | 1.84 | 5 | 6.99 | -0.03 |
| **$pH_{eq}$ = 7.33** | 10 | 9.45 | 2.43 | 10 | 6.71 | -0.62 |

Table S3: Concentrations of Co, Ni, ammonia $NH_4^+$ and phosphate $PO_4^{3-}$/total phosphorus (TP) in different wastewaters in mg/L, otherwise unit displayed; agricultural wastewaters contain high amounts of ammonia and phosphorus, while mine wastewaters exhibit high concentrations of heavy metals.

| country | region | waste water | species | concentration | reference |
|---|---|---|---|---|---|
| DRC | Katanga province | mine wastewater | Co | 3.164 | Atibu et al. 2013 [1] |
| Italy | Rio Piscinas | ground water (contaminated) | Co | 2.9-1.5 | Concas et al. 2006 [2] |
| | | | Ni | 4.6-3.0 | |
| Turkey | Amik plain | agricultural wastewater | $NH_4^+$ | 13 | Agca et al. 2014 [3] |
| | | | TP | 87 | |
| | | | Co | 13 [µg/L] | |
| | | | Ni | 50 [µg/L] | |
| USA | Duluth gabbro, Minnesota | mine wastewater | Co | 9.8 | Hammack et al. 1991 [4] |
| | | | Ni | 198 | |
| Jordan | Eshidiya mine | mine wastewater | $NH_4^+$ | 95-30 [µg/L] | Al-Hwaiti et al. 2016 [5] |
| | | | $PO_4^{3-}$ | 8.9-0.1 | |
| | | ground water | $NH_4^+$ | 3.8-3.5 | Al-Hwaiti et al. 2016 [5] |
| | | | $PO_4^{3-}$ | 50-10 [µg/L] | |
| China | Sichuan | agricultural wastewater | $NH_4^+$ | 1257.14 | Feng et al. 2020 [6] |
| | | | $PO_4^{3-}$ | 25.01 | |
| Brazil | | agricultural wastewater | $NH_4^+$ | 77 | Leite et al. 2009 [7] |
| | | | TP | 12 | |
| USA | Cobalt, Ontario | surface water | Co | 2028-0.5 [µg/L] | Percival et al. 1996 [8] |
| | | | Ni | 660-2.6 [µg/L] | |
| Serbia | Bor | Synthetic waste water (similar to mine waste waters) | Ni | 0.6128 | Stopic et al. 2007 [9] |
| | | | Co | 1.2 | |

Table S4: Pre-peak integration results by fitting Gaussian functions with all fitting parameters; Pre-peak fits visible in Figure S15.

| Prepeak area integration results | | | | | fit function: Gaussian | |
|---|---|---|---|---|---|---|
| Sample | chemical formula | $R^2$ value | Pre-peak Area $A_{pp}$ (err) | FWHM | Center | Max. Height |
| Co-octahydrate | $Co_3(PO_4)_2 \cdot 8H_2O$ | 0.995 | 0.0363(4) | 2.296 | 7711.2 | 0.0149 |
| Co-dittmarite | $NH_4CoPO_4 \cdot H_2O$ | 0.991 | 0.0355(4) | 2.101 | 7711.5 | 0.0159 |
| Co-struvite | $NH_4CoPO_4 \cdot 6H_2O$ | 0.993 | 0.0298(3) | 2.314 | 7711.1 | 0.0121 |
| Ni-struvite | $NH_4NiPO_4 \cdot 6H_2O$ | 0.992 | 0.0281(2) | 1.969 | 8333.3 | 0.0134 |

Table S5: ARTEMIS fit parameter for Ni-struvite R-factor= 0.005.

| sample | scattering path | degeneracy | $\sigma^2$ | $R_{diff}$ [Å] | $R_{diff}^2$ [Å²] | $R_{model}$ [Å] | $R_{fit}$ [Å] |
|---|---|---|---|---|---|---|---|
| Ni-struvite | Ni1-O1 | 4 | 0.009 | -0.001 | 1.00E-06 | 2.046 | **2.044** |
| NH$_4$NiPO$_4$•6H$_2$O | Ni1-O2 | 2 | 0.009 | -0.001 | 1.00E-06 | 2.089 | **2.088** |
| mean $R_{fit}$ (err) [Å] | Ni1-O3 H1 | 4 | -0.001 | -0.207 | 4.29E-02 | 2.560 | 2.353 |
| 2.059(6) | Ni1-H2 | 5 | 0.012 | -0.217 | 4.70E-02 | 2.586 | 2.369 |
| $V_{sphere}$ (err) [Å³] | Ni1-H3 | 2 | 0.012 | -0.217 | 4.70E-02 | 2.627 | 2.410 |
| 36.6 (2) | Ni1-H4 | 2 | 0.012 | -0.217 | 4.70E-02 | 2.668 | 2.451 |
| R-factor | Ni1-O1-O1 | 6 | -0.001 | -0.207 | 4.29E-02 | 3.465 | 3.258 |
| 0.005 | Ni1-O3-O2 | 16 | -0.001 | -0.207 | 4.29E-02 | 3.530 | 3.322 |
| $S_0^2$ Amplitude reduction factor (err) | Ni1-N1 | 1 | 0.005 | -0.186 | 3.47E-02 | 4.022 | 3.836 |
| 1.3(1) | Ni1-N2 | 2 | 0.005 | -0.186 | 3.47E-02 | 4.111 | 3.925 |
| ΔE | Ni1-O4 | 4 | 0.006 | 0.242 | 5.85E-02 | 4.084 | 4.326 |
| 3.1(6) | Ni1-O5 | 2 | 0.006 | 0.242 | 5.85E-02 | 4.162 | 4.404 |
| | Ni1-O6 | 2 | 0.006 | 0.242 | 5.85E-02 | 4.200 | 4.442 |
| | Ni1-O7 | 3 | 0.006 | 0.242 | 5.85E-02 | 4.272 | 4.514 |
| | Ni1-O8 | 2 | 0.006 | 0.242 | 5.85E-02 | 4.329 | 4.571 |

Table S6: ARTEMIS fit parameter for Co-struvite R-factor= 0.012.

| sample | scattering path | degeneracy | $\sigma^2$ | $R_{diff}$ [Å] | $R_{diff}^2$ [Å$^2$] | $R_{model}$ [Å] | $R_{fit}$ [Å] |
|---|---|---|---|---|---|---|---|
| Co-struvite | Co1-O1 | 2 | 0.005 | 0.0001 | 1.69E-08 | 2.058 | **2.058** |
| NH$_4$CoPO$_4$•6H$_2$O | Co1-O2 | 2 | 0.005 | 0.0001 | 1.69E-08 | 2.089 | **2.089** |
| mean $R_{fit}$ (err) [Å] | Co1-O3 | 2 | 0.005 | 0.0001 | 1.69E-08 | 2.129 | **2.129** |
| **2.091(1)** | Co1-H1 | 2 | 0.003 | 0.3078 | 9.48E-02 | 2.211 | 2.519 |
| $V_{sphere}$ (err) [Å$^3$] | Co1-H2 | 2 | 0.003 | 0.3078 | 9.48E-02 | 2.451 | 2.759 |
| **38.35(2)** | Co1-H3 | 4 | 0.003 | 0.3078 | 9.48E-02 | 2.549 | 2.857 |
| R-factor | Co1-H4 | 2 | 0.003 | 0.3078 | 9.48E-02 | 2.616 | 2.924 |
| **0.012** | Co1-O1-H3-O1 | 4 | 0.003 | 0.3774 | 1.42E-01 | 2.752 | 3.129 |
| $S_0^2$ Amplitude reduction factor (err) | Co1-O5-O6-O5 | 2 | 0.003 | 0.3774 | 1.42E-01 | 2.840 | 3.217 |
| **0.85(5)** | Co1-O3-H5-O3 | 1 | 0.003 | 0.3774 | 1.42E-01 | 2.843 | 3.220 |
| ΔE | Co1-O1-O2 | 6 | -0.003 | 0.2301 | 5.29E-02 | 3.507 | 3.737 |
| **4.5(6)** | Co1-O1-O3 | 12 | -0.003 | 0.2301 | 5.29E-02 | 3.575 | 3.805 |
| | Co1-O2-O3 | 4 | -0.003 | 0.2301 | 5.29E-02 | 3.622 | 3.852 |
| | Co1-N1 | 1 | 0.001 | -0.0171 | 2.94E-04 | 4.027 | 4.009 |
| | Co1-N2 | 2 | 0.001 | -0.0171 | 2.94E-04 | 4.133 | 4.115 |
| | Co1-O4 | 4 | 0.003 | 0.3774 | 1.42E-01 | 4.105 | 4.482 |
| | Co1-O6 | 4 | 0.003 | 0.3774 | 1.42E-01 | 4.199 | 4.576 |
| | Co1-O7 | 3 | 0.003 | 0.3774 | 1.42E-01 | 4.288 | 4.665 |
| | Co1-O8 | 2 | 0.003 | 0.3774 | 1.42E-01 | 4.362 | 4.739 |
| | Co1-P1 | 2 | 0.003 | 0.3774 | 1.42E-01 | 4.393 | 4.770 |

Table S7: ARTEMIS fit parameter for Co-dittmarite (COD), R-factor= 0.008.

| sample | scattering path | degeneracy | $\sigma^2$ | $R_{diff}$ [Å] | $R_{diff}^2$ [Å$^2$] | $R_{model}$ [Å] | $R_{fit}$ [Å] |
|---|---|---|---|---|---|---|---|
| **Co-dittmarite** | Co1-O1 | 3 | 0.001 | 0.0001 | 1.69E-08 | 2.055 | **2.023** |
| **NH$_4$CoPO$_4$•H$_2$O** | Co1-O2 | 3 | 0.001 | 0.0001 | 1.69E-08 | 2.201 | **2.169** |
| **mean $R_{fit}$ (err) [Å]** | Co1-H1 | 2 | -0.008 | 0.0001 | 1.69E-08 | 2.492 | 2.222 |
| **2.096(1)** | Co1-H2 | 2 | -0.008 | 0.3078 | 9.48E-02 | 2.970 | 2.699 |
| **$V_{sphere}$ (err)[Å$^3$]** | Co1-P1 | 1 | 0.002 | 0.3078 | 9.48E-02 | 2.797 | 2.797 |
| **38.55(2)** | Co1-P2 | 2 | 0.002 | 0.3078 | 9.48E-02 | 3.209 | 3.209 |
| **R-factor** | Co1-O3 | 2 | 0.001 | 0.2301 | 5.29E-02 | 3.313 | 3.281 |
| **0.008** | Co1-P3 | 1 | 0.002 | 0.3078 | 9.48E-02 | 3.305 | 3.305 |
| **$S_0^2$ Amplitude reduction factor** | Co1-O4 | 2 | 0.001 | -0.0171 | 2.94E-04 | 3.592 | 3.560 |
| **0.79(6)** | Co1-Co2 | 7 | 0.010 | 0.3774 | 1.42E-01 | 3.712 | 3.653 |
| **ΔE** | Co1-O5 | 1 | 0.001 | -0.0171 | 2.94E-04 | 3.753 | 3.721 |
| **2.6(6)** | Co1-N1 | 1 | -0.008 | 0.3774 | 1.42E-01 | 4.010 | 3.740 |
| | Co1-O6 | 3 | 0.001 | 0.2301 | 5.29E-02 | 3.813 | 3.781 |
| | Co1-O7 | 1 | 0.001 | 0.2301 | 5.29E-02 | 3.944 | 3.912 |

Table S8: ARTEMIS fit parameter for Co-phosphate octahydrate (CPO), R-factor= 0.005.

| sample | scattering path | degeneracy | $\sigma^2$ | $R_{diff}$ [Å] | $R_{diff}^2$ [Å$^2$] | $R_{model}$ [Å] | $R_{fit}$ [Å] |
|---|---|---|---|---|---|---|---|
| **Co-octahydrate** | Co1-O1 | 5 | 0.011 | -0.003 | 8.53E-06 | 2.021 | **2.099** |
| **Co$_3$(PO$_4$)$_2$•8H$_2$O** | Co1-O2 | 1 | 0.011 | -0.003 | 8.53E-06 | 2.158 | **2.155** |
| **mean $R_{fit}$ (err) [Å]** | Co1-H1 | 16 | 0.018 | -0.079 | 6.19E-03 | 2.451 | 2.372 |
| **2.11(2)** | Co1-O2-H2 | 8 | 0.001 | 0.416 | 1.73E-01 | 2.593 | 2.593 |
| **V$_{sphere}$ (err)[Å$^3$]** | Co1-H3 | 4 | 0.018 | -0.079 | 6.19E-03 | 3.226 | 3.147 |
| **39.3(6)** | Co1-P1 | 2 | 0.010 | -0.003 | 9.18E-06 | 3.194 | 3.191 |
| **R-factor** | Co1-H3 | 8 | 0.018 | -0.079 | 6.19E-03 | 3.319 | 3.240 |
| **0.005** | Co1-O3 | 4 | 0.001 | 0.416 | 1.73E-01 | 3.579 | 3.389 |
| **S$_0^2$ Amplitude reduction factor (err)** | Co1-O1-O2 | 8 | -0.013 | 0.099 | 9.82E-03 | 3.553 | 3.553 |
| **1.3(1)** | Co1-O5 | 2 | 0.010 | -0.003 | 9.18E-06 | 3.675 | 3.576 |
| **ΔE** | Co1-O1-O2 | 8 | -0.013 | 0.099 | 9.82E-03 | 3.584 | 3.584 |
| **4.9(7)** | Co1-O2-O2 | 8 | 0.010 | -0.003 | 9.18E-06 | 3.684 | 3.672 |
| | Co1-O1-P1 | 4 | -0.013 | 0.099 | 9.82E-03 | 3.389 | 3.684 |
| | Co1-O4 | 8 | 0.010 | -0.003 | 9.18E-06 | 3.950 | 3.947 |
| | Co1-H3-O4 | 16 | 0.001 | 0.416 | 1.73E-01 | 4.014 | 4.014 |
| | Co1-O6-H5 | 8 | 0.001 | 0.416 | 1.73E-01 | 4.450 | 4.450 |

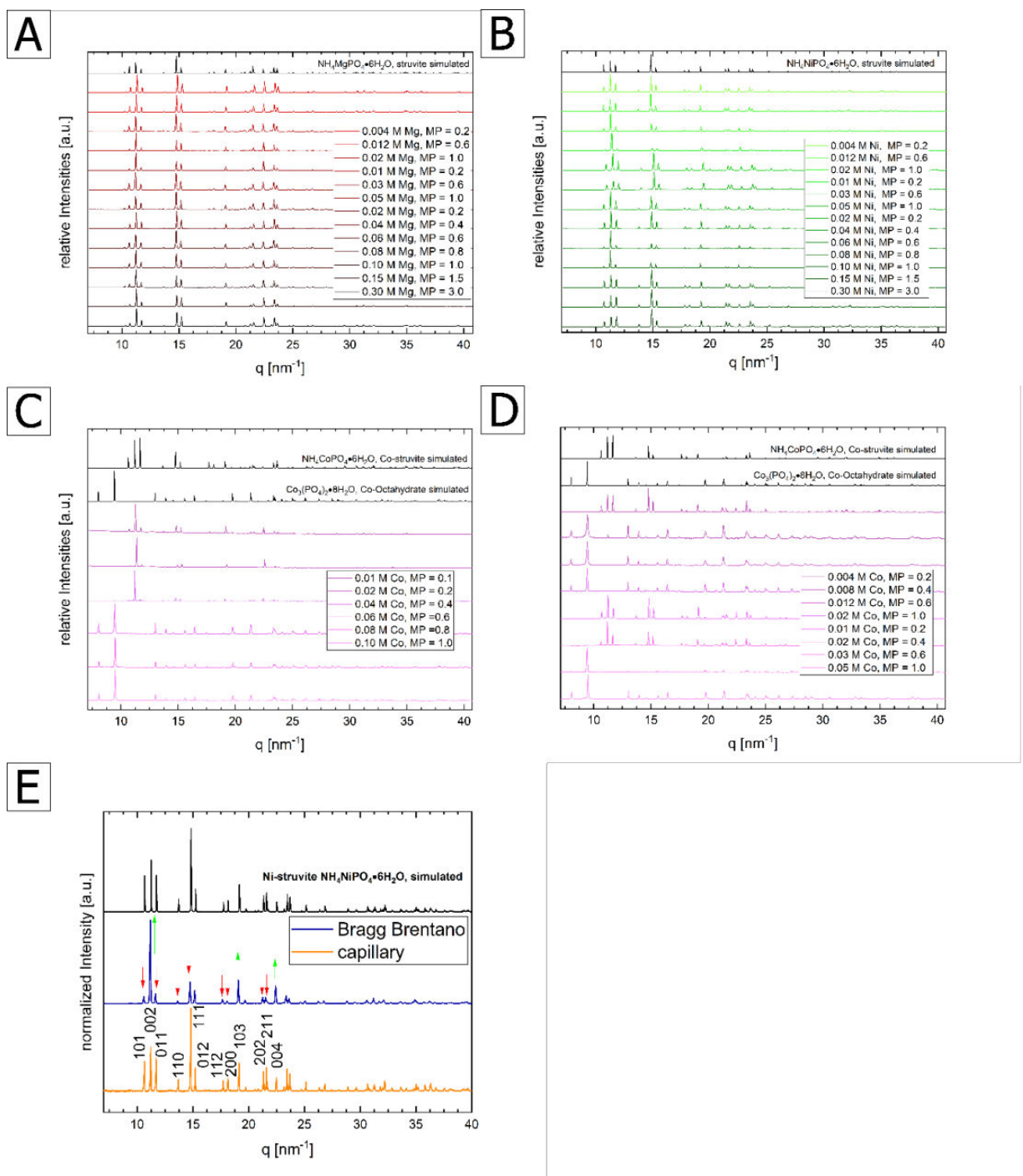

Figure S1: Stacked XRD patterns with the simulated pattern of the phosphate compound in black; (A) Mg-series; (B) Ni-series; (C) and (D) Co-series for different concentrations, according to the legends; These XRD patterns were used to create the phase stability diagrams; Mg-struvite reference COD 9007674; Ni-struvite reference ICSD 403058; Co-struvite reference ICSD 170042; Co(II) phosphate octahydrate reference COD 2020362; (E) In all measured XRD patterns of M-struvite preferred orientation of the (001) crystal planes occurred to a varying degree, and was visible through strong intensity discrepancies between the simulated and measured patterns. All crystal planes with high (h00) or (0k0) component exhibit lower intensities while reflexes with a dominant (00l) component are intensified than in the simulated pattern visible in (E). The preferred orientation is caused by the alignment of elongated habit of the crystals during the preparation of the flat powder sample holders.

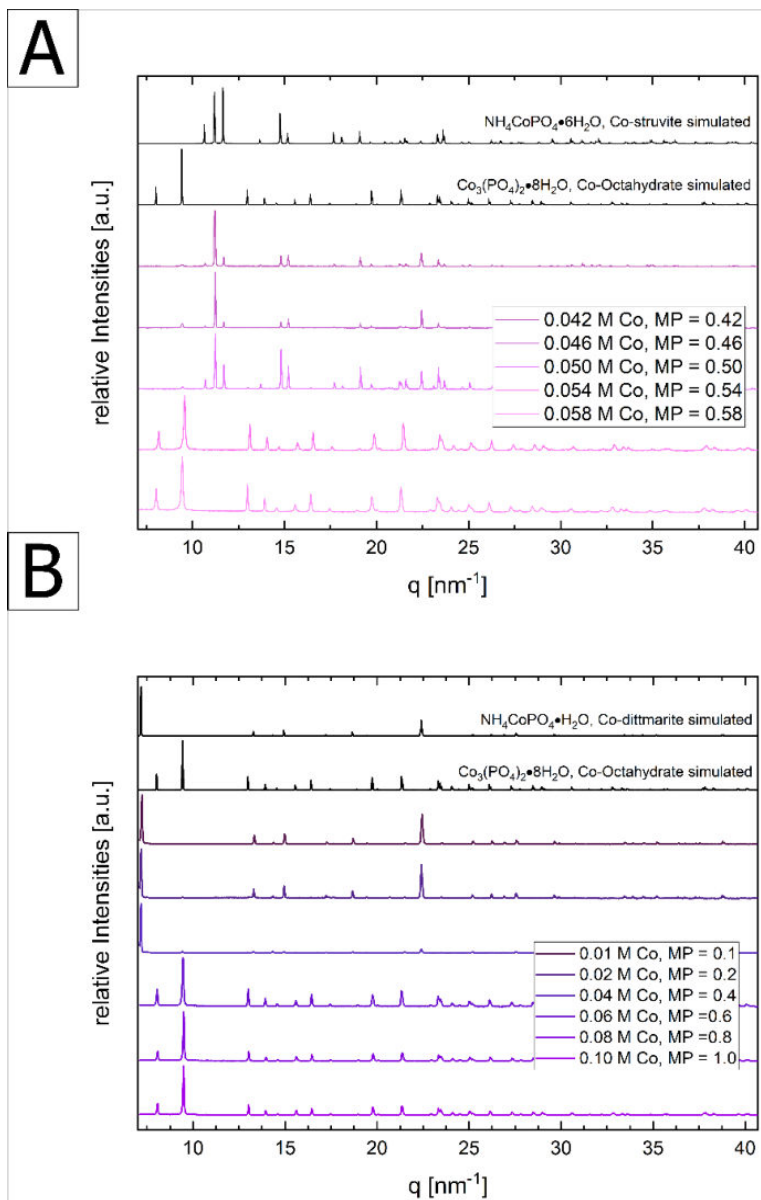

Figure S2: Stacked XRD patterns with the simulated pattern of the phosphate compound in black; (A) In the transition interval between M/P ratios of 0.58-0.42 with 0.1 M DAP of the Co-system Co-struvite and Co(II) phosphate octahydrate are stable; (B) Co-struvite decomposes in air to Co-dittmarite, while Co(II)phosphate octahydrate remains stable under these conditions. The same simulated patterns used as in Figure S2 with added Co-dittmarite COD 2008122.

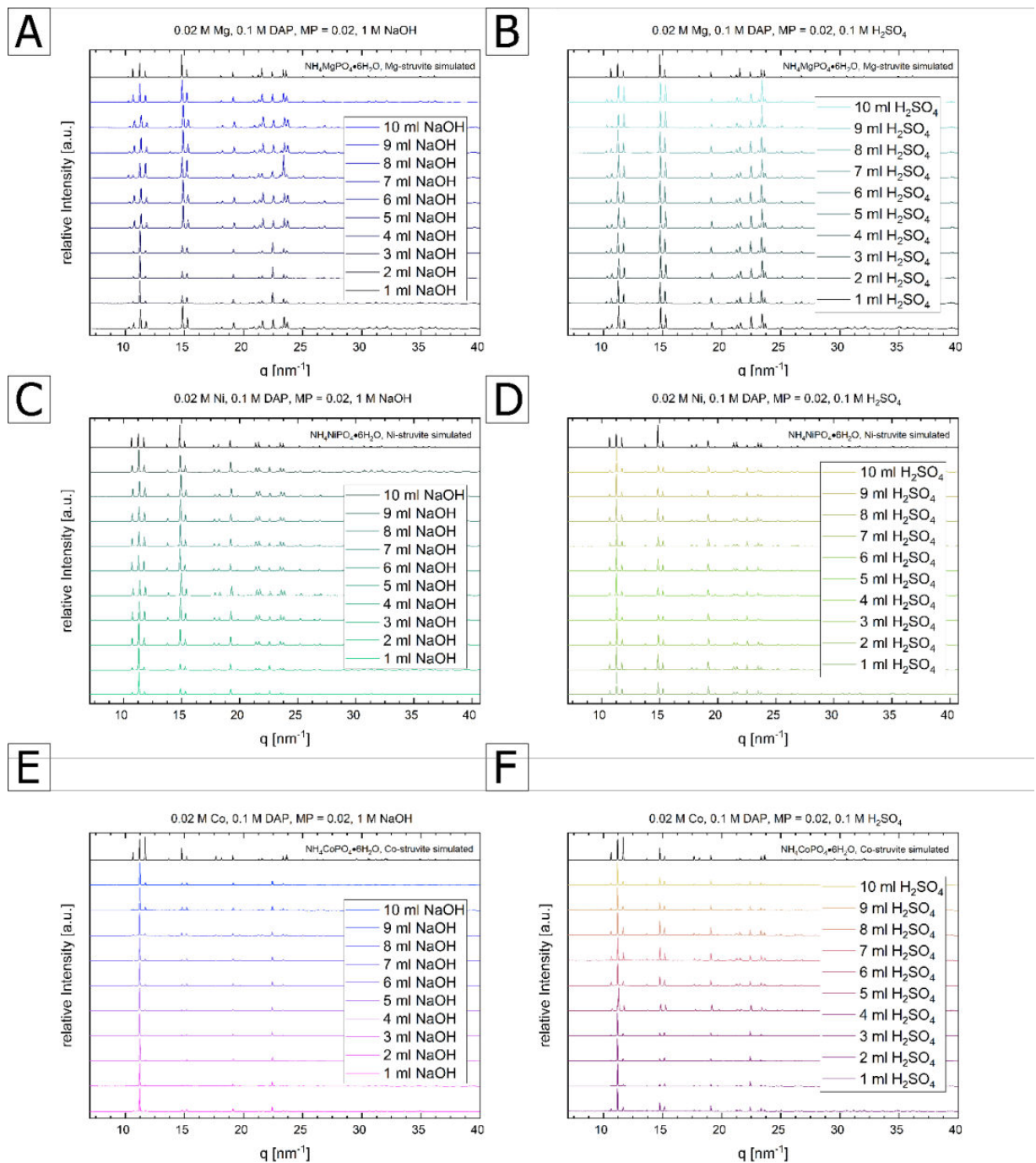

Figure S3: pH adjusted samples with NaOH$_{(aq)}$ and H$_2$SO$_{4(aq)}$ of (A) (B) Mg; of (C) (D) Ni and (E) (F) Co with their respective experimental conditions (see legends) and their phase composition. In all samples single phase M-struvite precipitated; same simulated patterns used as in Figure S2.

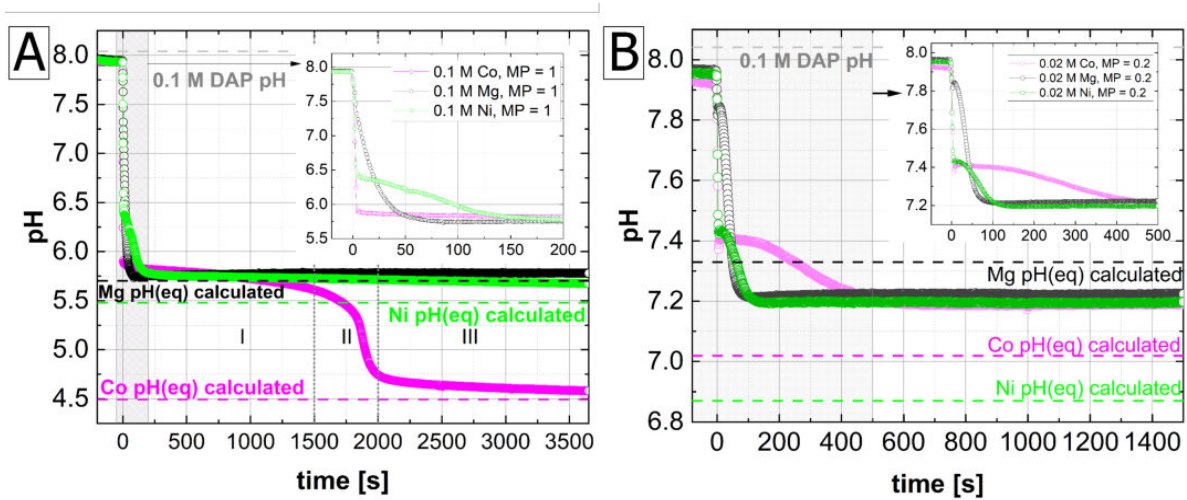

Figure S4: Time-resolved pH measurements with 1 s resolution (A) for the high concentration samples and (B) the low concentration samples; stage I, II, III indicate different slopes in the pH curve of the 0.1 M Co sample; insets show in detail the pH evolution over the first 200 s; the axes and units in the insets are the same as in the main graphs.

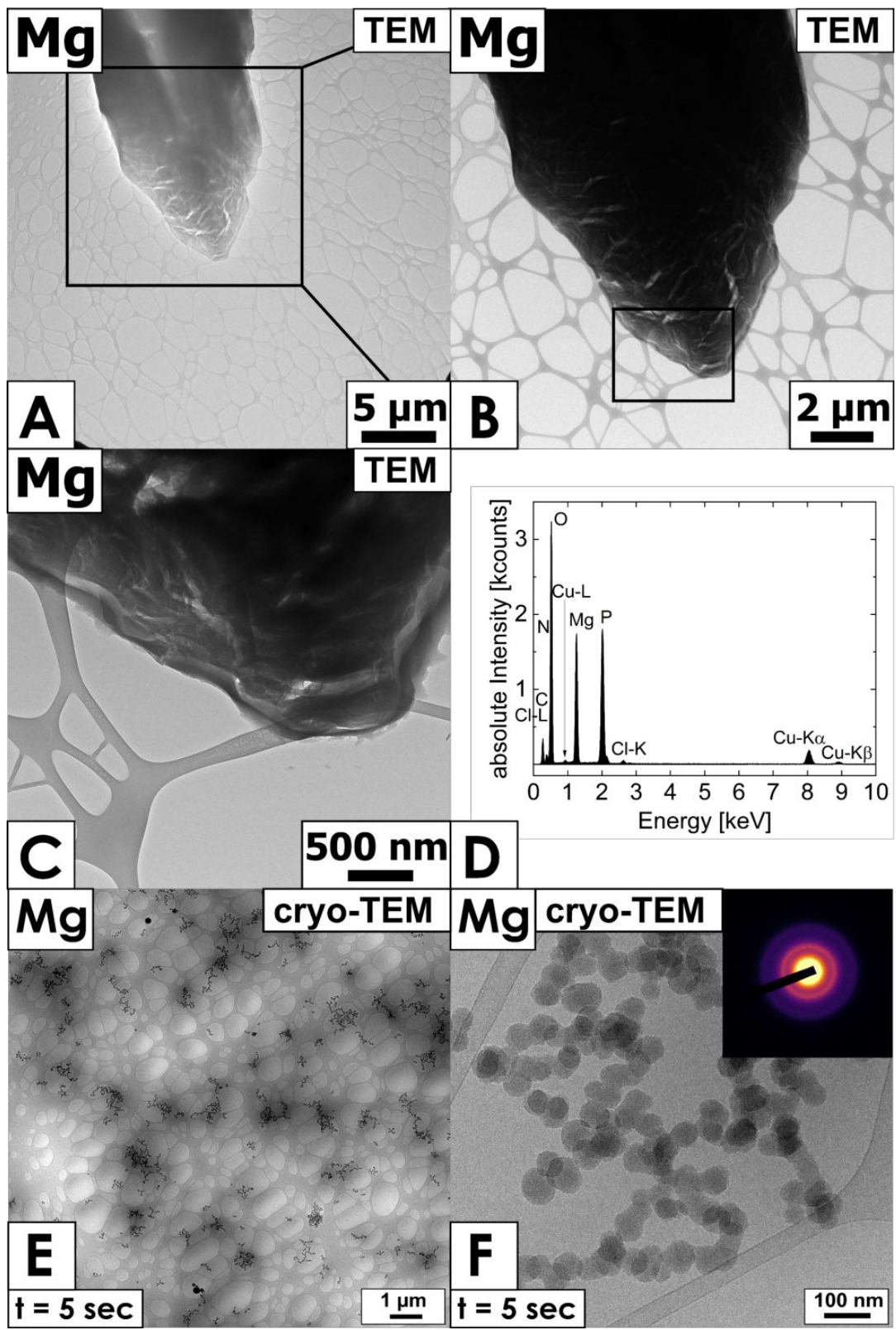

Figure S5: (A) Dry TEM images of Mg sample after 5 seconds mixing forming crystals of µm-size with visible cracks; (B) detail view of Mg-struvite crystal from the area marked with a black rectangle in (A); (C) close view of the Mg-struvite crystal from the area marked with a black rectangle in (B); (D) EDS spectrum of Mg-struvite; (E) cryo-TEM image overview of amorphous nanophases with t = 5 seconds mixing time; (F) cryo-TEM image with detailed view of amorphous nanophases in the Mg-sample and associated SAED pattern; Note the difference of the observed structures between dry- and cryo-TEM.

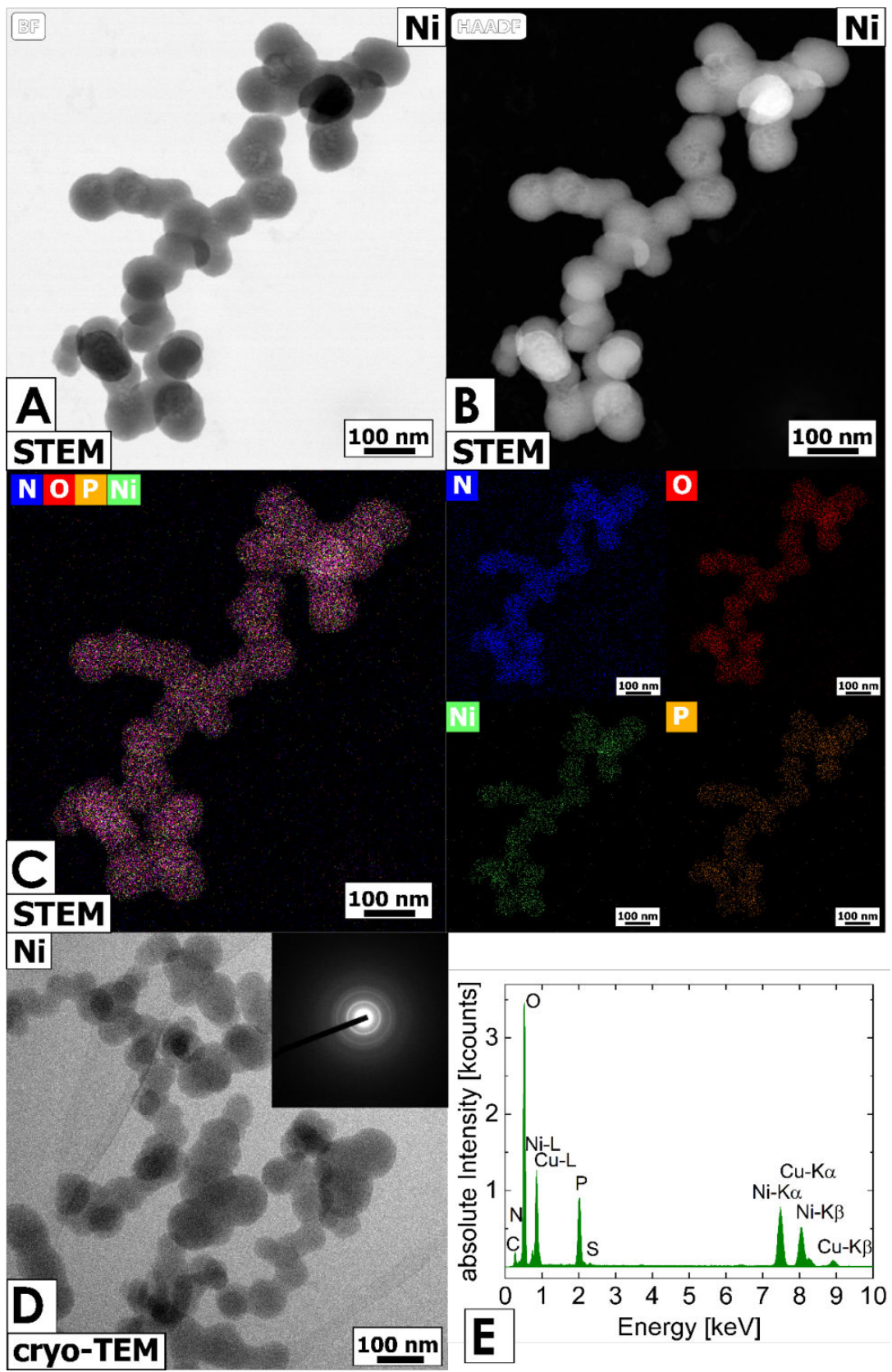

Figure S6: (A) Dry STEM analysis of Ni sample after t = 5 sec mixing with the formation of amorphous nanophases in bright field = BF; (B) high angular dark field = HAADF; (C) and elemental mappings of N = nitrogen, O = oxygen, Ni = nickel and P = phosphorus; (D) cryo-TEM image of amorphous Ni-phase with the SAED pattern in the inset; (E) EDS point measurement from amorphous particles.

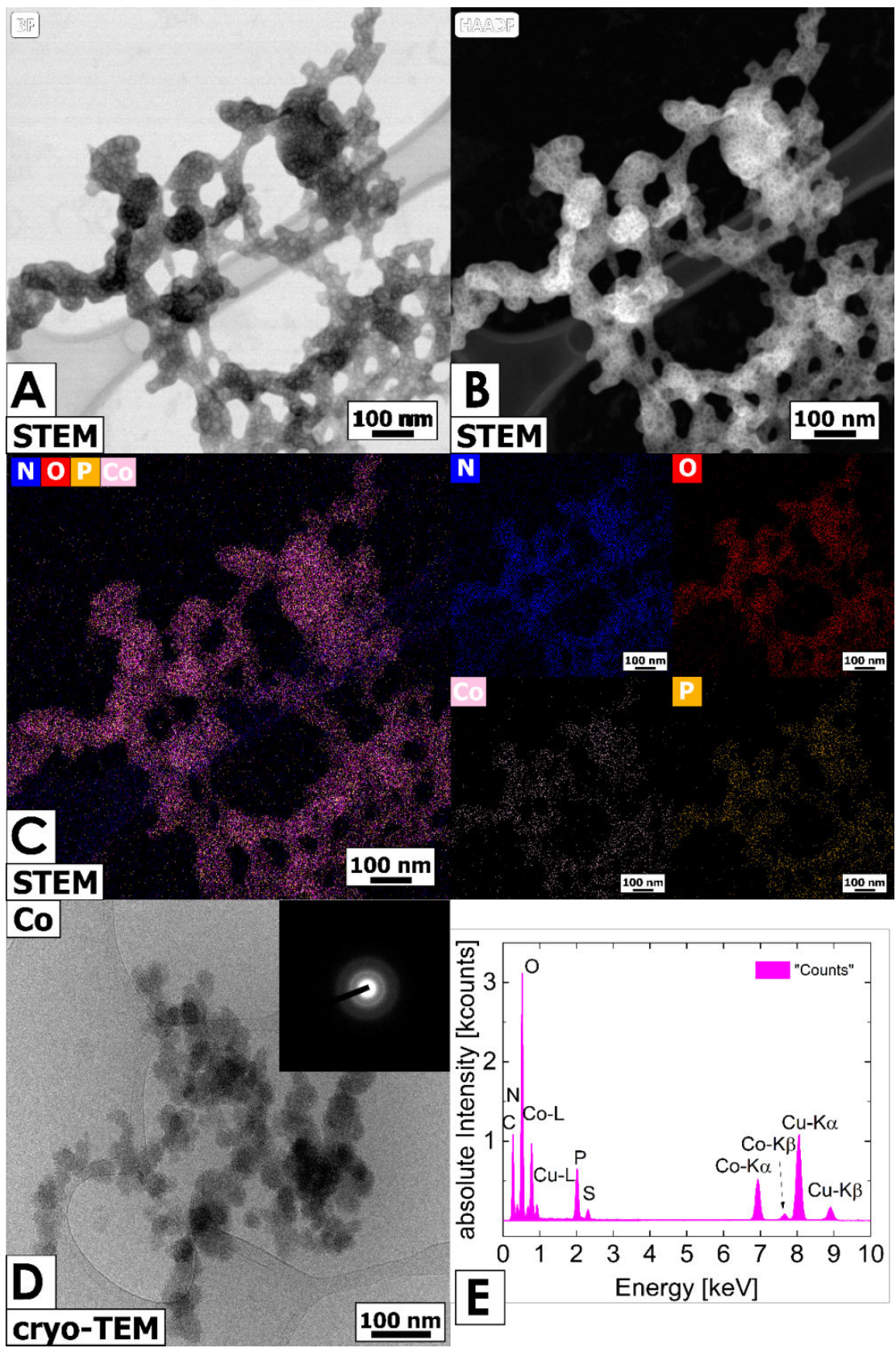

Figure S7: (A) Dry STEM analysis of Co sample after t = 5 sec mixing; formation of amorphous nanoparticles in bright field = BF; (B) high angular dark field = HAADF; (C) elemental mappings of N = nitrogen, O = oxygen, Co = cobalt and P = phosphorus; (D) cryo-TEM image of amorphous Co-phase with the SAED pattern; (E) EDS point measurement of amorphous particles.

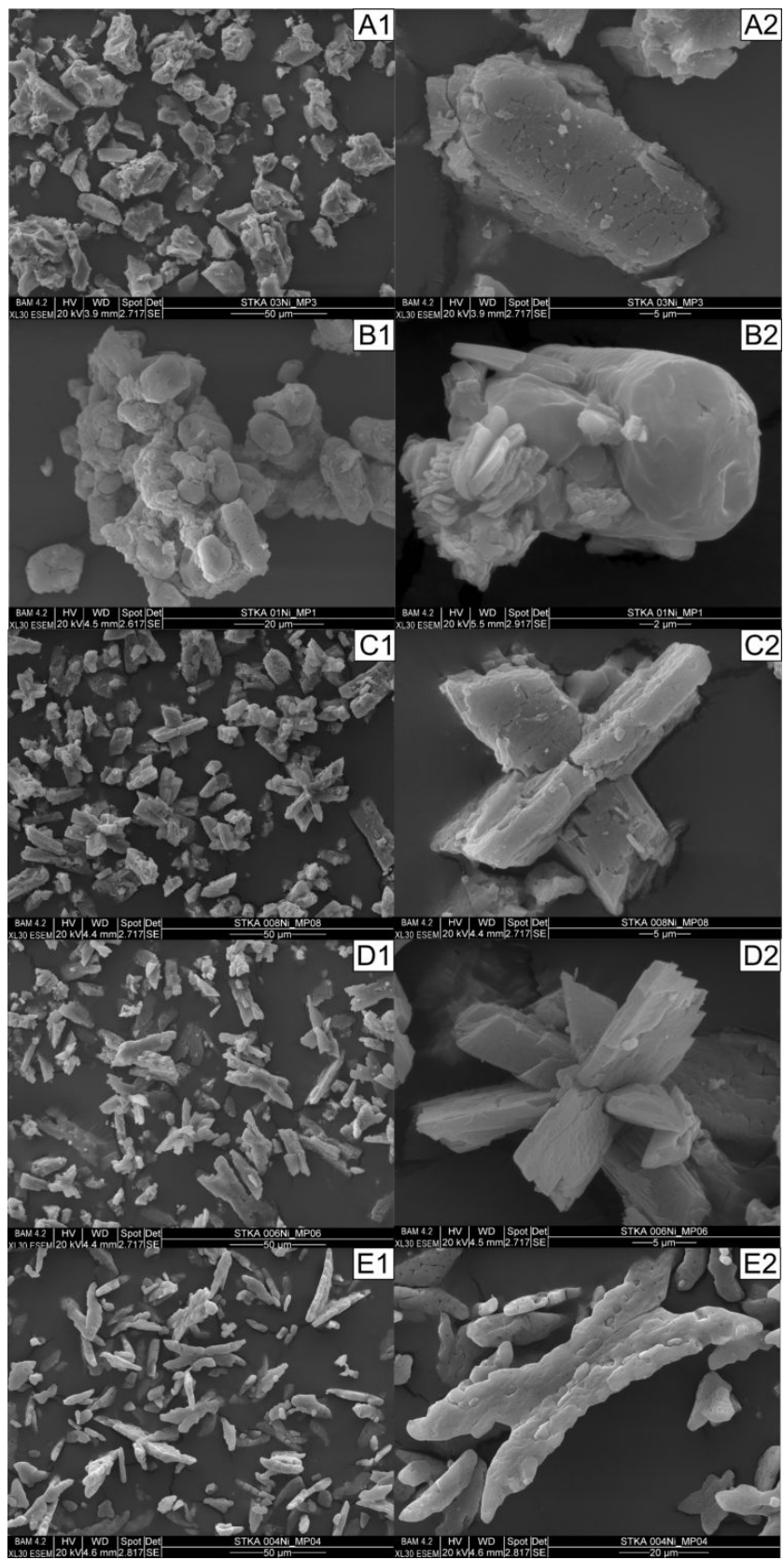

Figure S8: SE images of Ni-struvite synthesized at varying concentrations of the reactants; on the left side: an overview image at a low magnification marked always by 1, on the right a detailed view at q crystal from the respective sample marked with a 2; (A) 0.3 M Ni, 0.1 M DAP, M/P ratio = 3; (B) 0.1 M Ni, 0.1 M DAP, M/P ratio = 1; (C) 0.08 M Ni, 0.1 M DAP, M/P ratio = 0.8; (D) 0.06 M Ni, 0.1 M DAP, M/P ratio = 0.6; (E) 0.04 M Ni, 0.1M DAP, M/P ratio = 0.4;

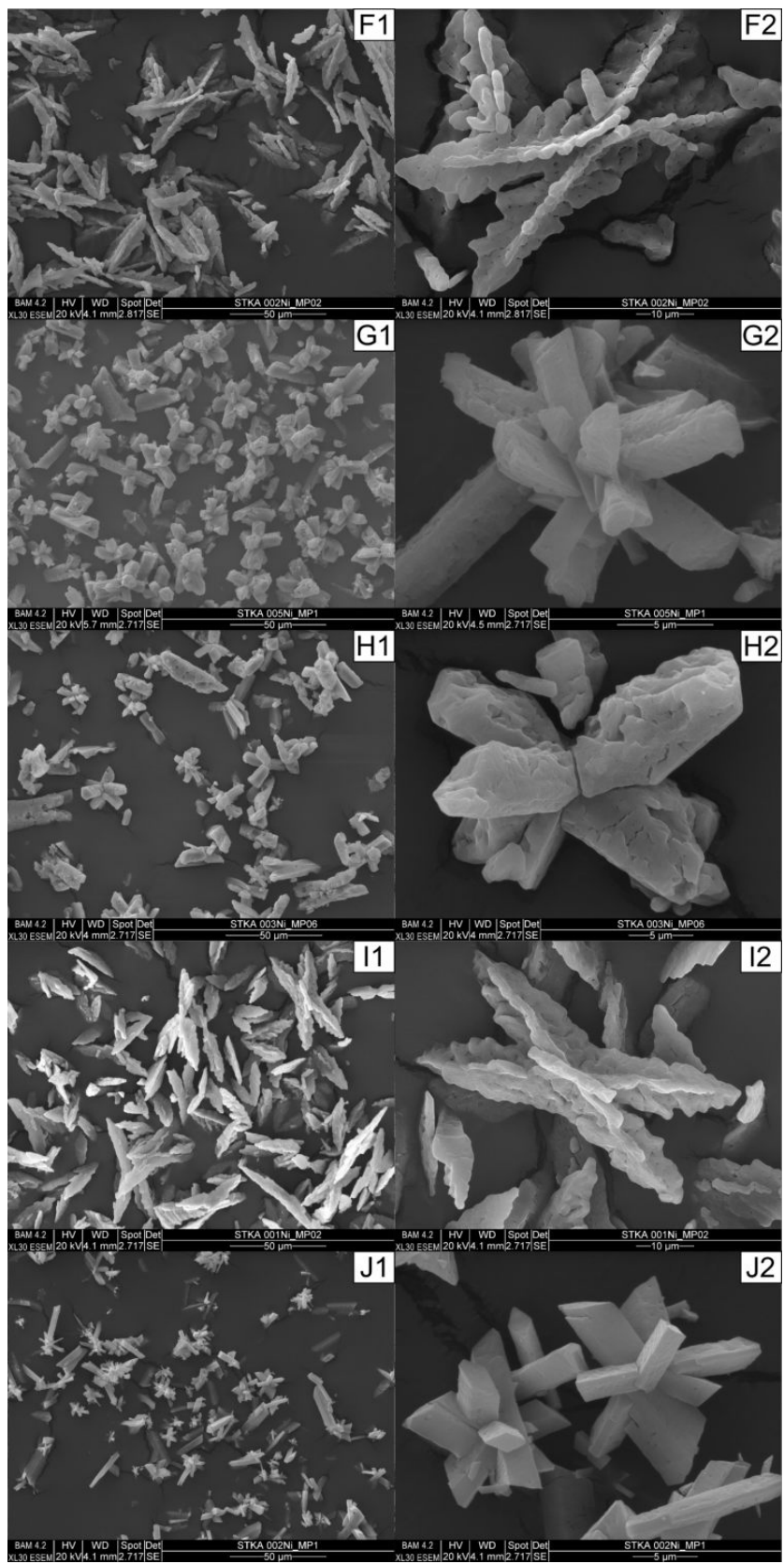

Figure S9: SE images of Ni-struvite synthesized at varying concentrations of the reactants; on the left side: an overview image at a low magnification marked always by 1, on the right a detailed view at a crystal from the respective sample marked with a 2; (F) 0.02 M Ni, 0.1 M DAP, M/P ratio = 0.2; (G) 0.05 M Ni, 0.05 M DAP, M/P ratio = 1.0; (H) 0.03 M Ni, 0.05 M DAP, M/P ratio = 0.6; (I) 0.01 M Ni, 0.05M DAP, M/P ratio = 0.2; (J) 0.02 M Ni, 0.02 M DAP, M/P ratio = 1.0.

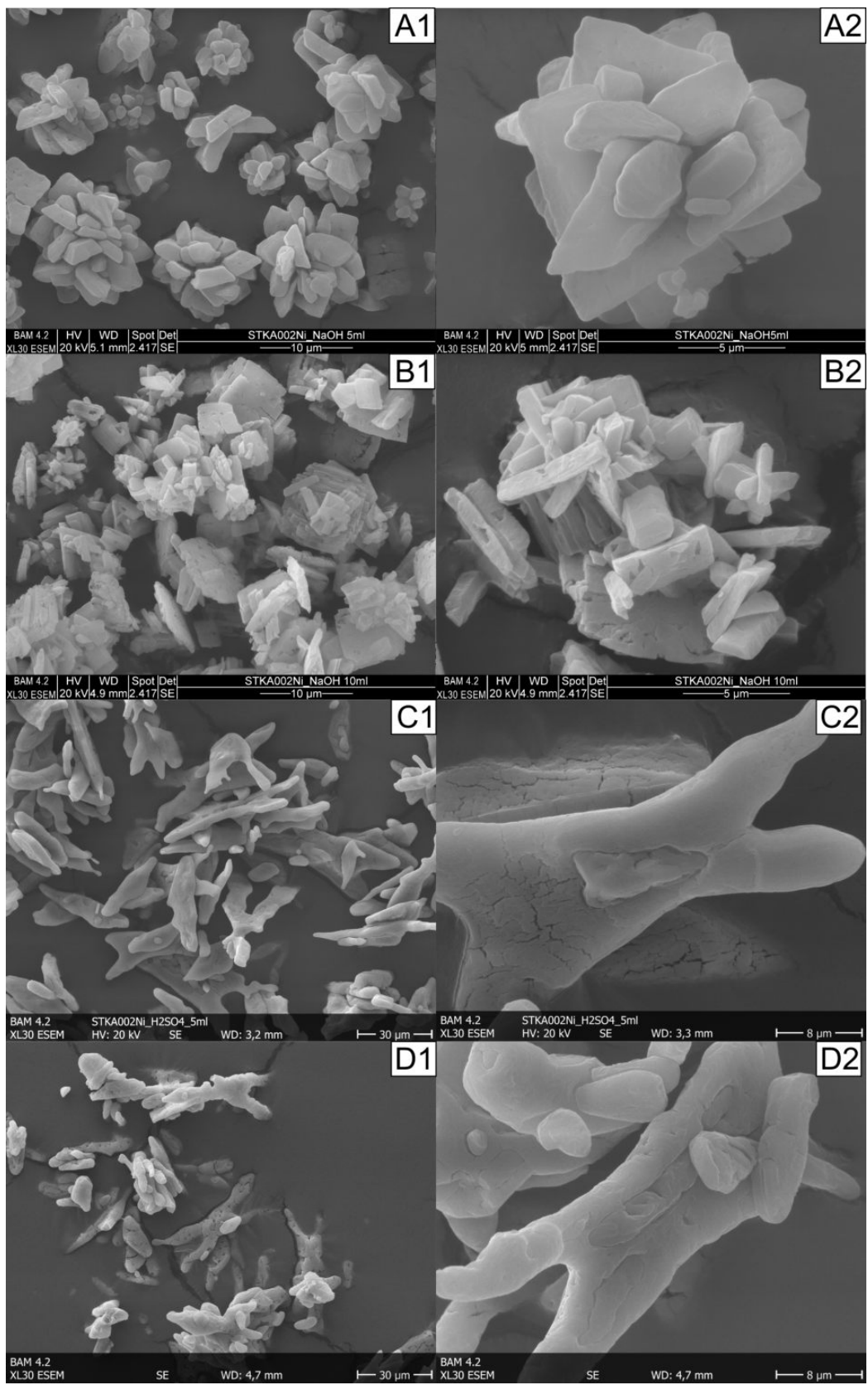

Figure S10: SE images of Ni-struvite synthesized from 0.02 M Ni and 0.1 M DAP leading to a M/P ratio of 0.2 at different initial pH with 5 and 10 ml 1 M NaOH$_{(aq)}$ or 0.1 M H$_2$SO$_{4(aq)}$; overview image marked by 1, detailed view at a selected crystal marked by 2; (A) 5 ml 1 M NaOH$_{(aq)}$; (B) 10 ml 1 M NaOH$_{(aq)}$; (C) 5 ml 1 M H$_2$SO$_{4(aq)}$; (D) 10 ml 1 M H$_2$SO$_{4(aq)}$.

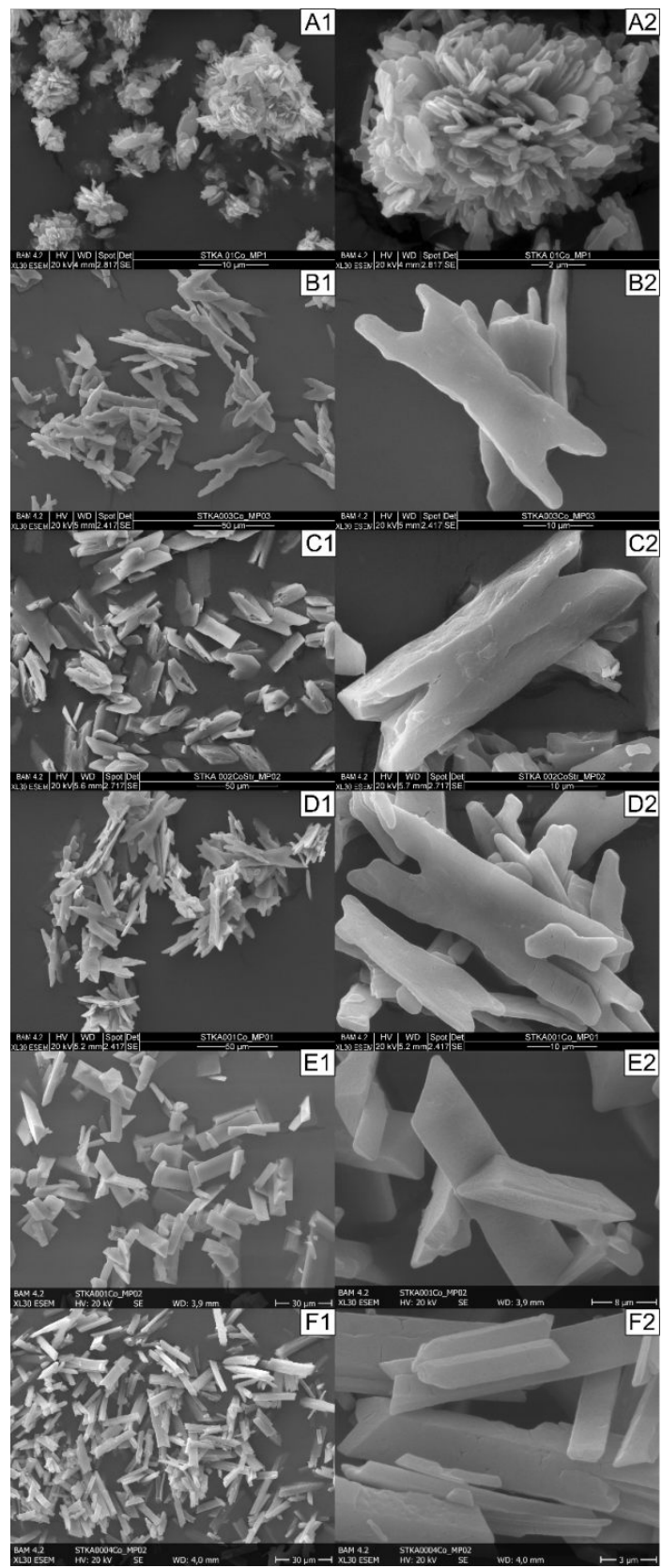

Figure S11: SE images of Co-phosphates synthesized at different concentrations; overview image marked by 1, detailed view at a selected crystal marked by 2; (A) Co(II)phosphate octahydrate 0.1 M Co, 0.1 M DAP, M/P ratio = 1; (B) 0.03 M Co, 0.1 M DAP, M/P ratio = 0.3; (C) 0.02 M Co, 0.1 M DAP, M/P ratio = 0.2; (D) 0.01 M Co, 0.1 M DAP, M/P ratio = 0.1; (E) 0.01 M Co, 0.05 M DAP, M/P ratio = 0.2; (F) 0.004 M Co, 0.02 M DAP, M/P ratio = 0.2;

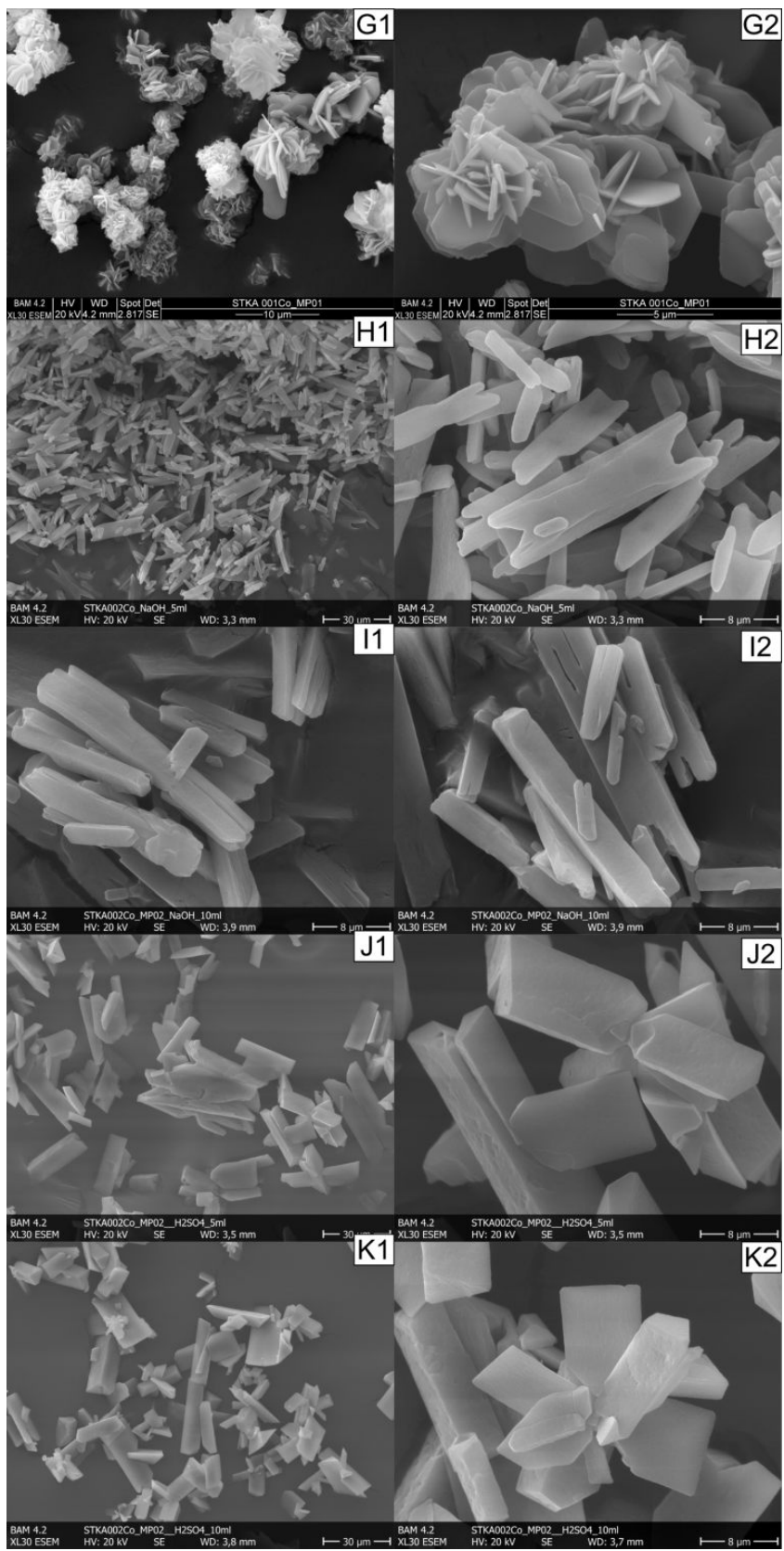

Figure S12: Co-struvite synthesized from 0.02 M Co and 0.1 M DAP leading to a M/P ratio of 0.2 at different initial pH with 5 ml and 10 ml 1 M NaOH$_{(aq)}$ or 0.1M H$_2$SO$_{4(aq)}$; overview image marked by 1, detailed view at a selected crystal marked by 2; (A) 5 ml 1 M NaOH$_{(aq)}$; (B) 10 ml 1 M NaOH$_{(aq)}$; (C) 5 ml 1 M H$_2$SO$_{4(aq)}$; (D) 10 ml 1 M H$_2$SO$_{4(aq)}$.

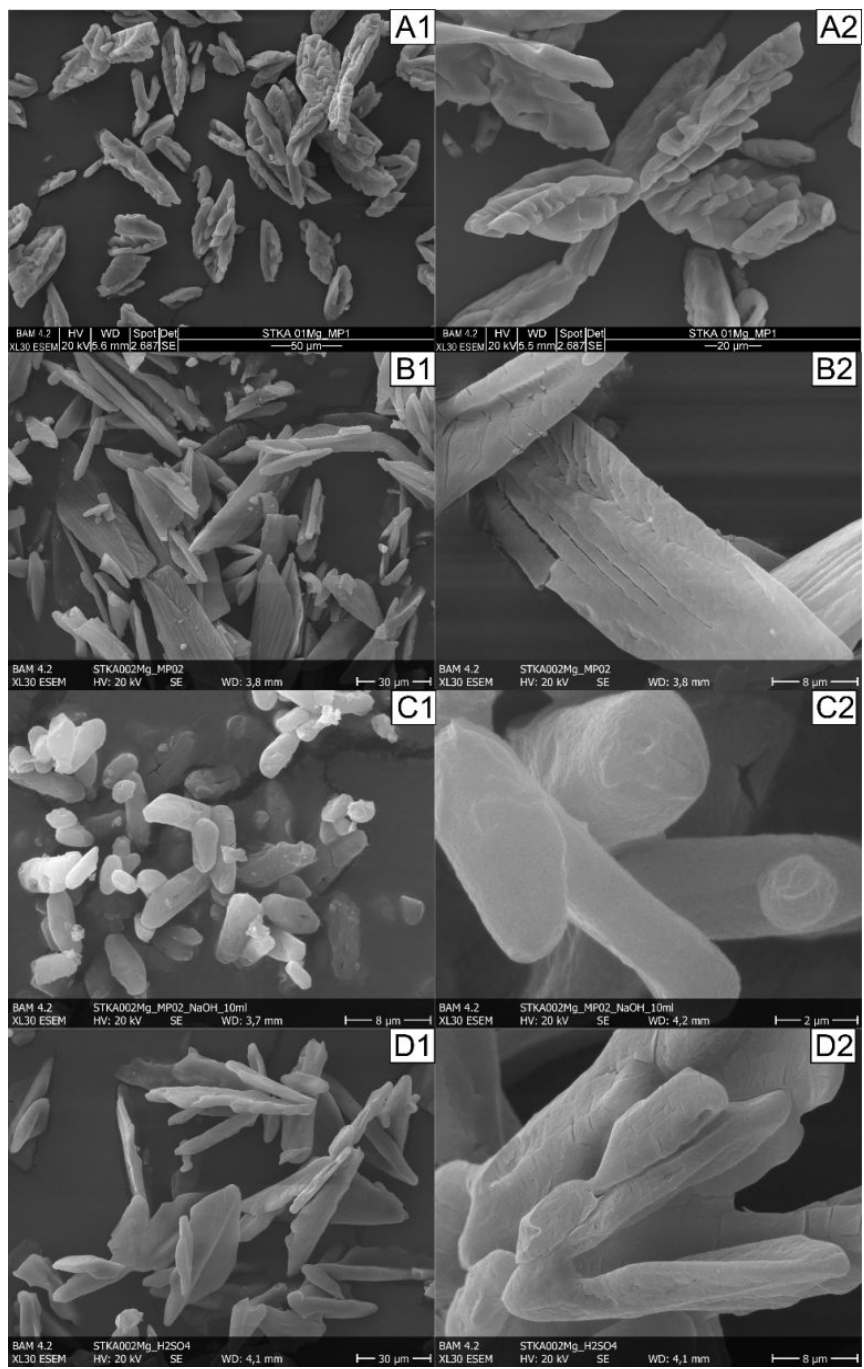

Figure S13: SE images of different Mg-struvites at low M/P ratios of 0.2 with pH adjusted samples at the sides and one sample with high M/P ratio of 1; overview image marked by 1, detailed view at a selected crystal marked by 2; (A) 0.1 M Mg, 0.1 M DAP; (B) 0.02 M Mg, 0.1 M DAP; (C) 0.02 M Mg, 0.1 M DAP and 10 ml 1 M NaOH$_{(aq)}$; (D) 0.02 M Mg, 0.1 M DAP and 10 ml 1 M H$_2$SO$_{4(aq)}$.

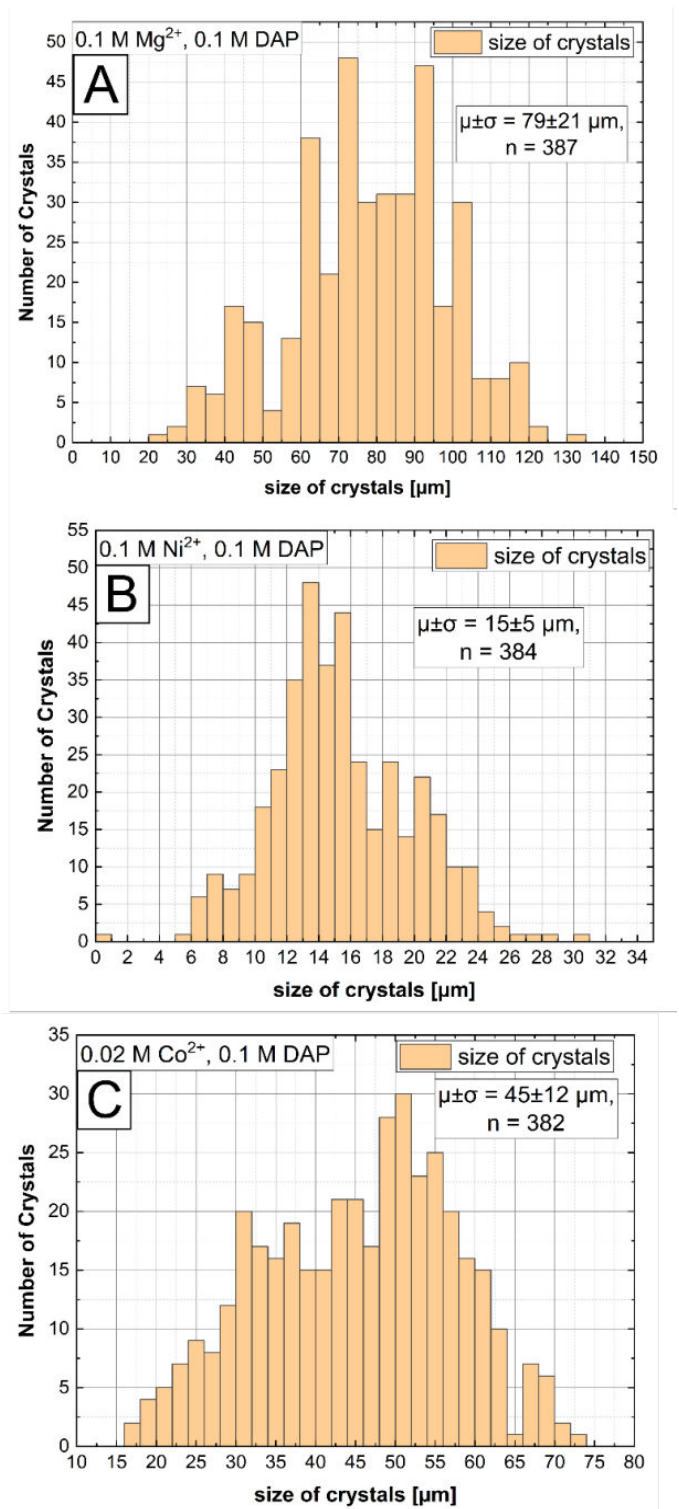

Figure S14: Exemplary size distribution histograms from selected M-struvite samples of M = Mg, Ni and Co; (A) histogram of 0.1 M Mg, 0.1 M DAP with a total sample population of 387, µ ± σ = 79 ± 21; (B) histogram of 0.1 M Ni, 0.1 M DAP with a total sample population of 384, µ ± σ = 15 ± 5; (C) histogram of 0.02 M Co, 0.1 M DAP with a total sample population of 384, µ ± σ = 15 ± 5 µm.

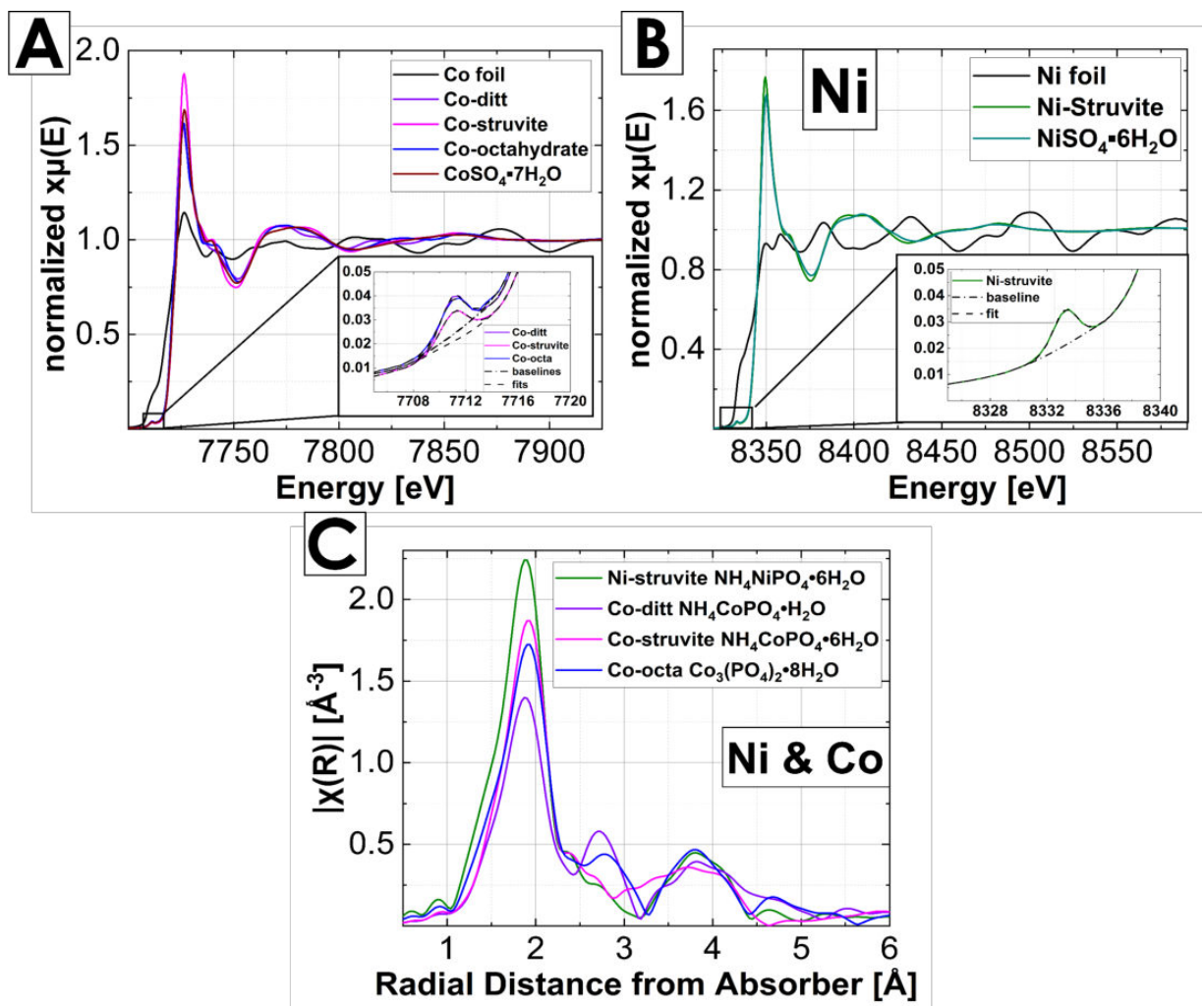

Figure S15: XAS spectra of (A) Co and (B) Ni samples with detailed view of the pre-peak region; (C) Fourier-transformed Ni- and Co spectra plotted in R-space; the axes and units in the insets are the same as in the main graphs.

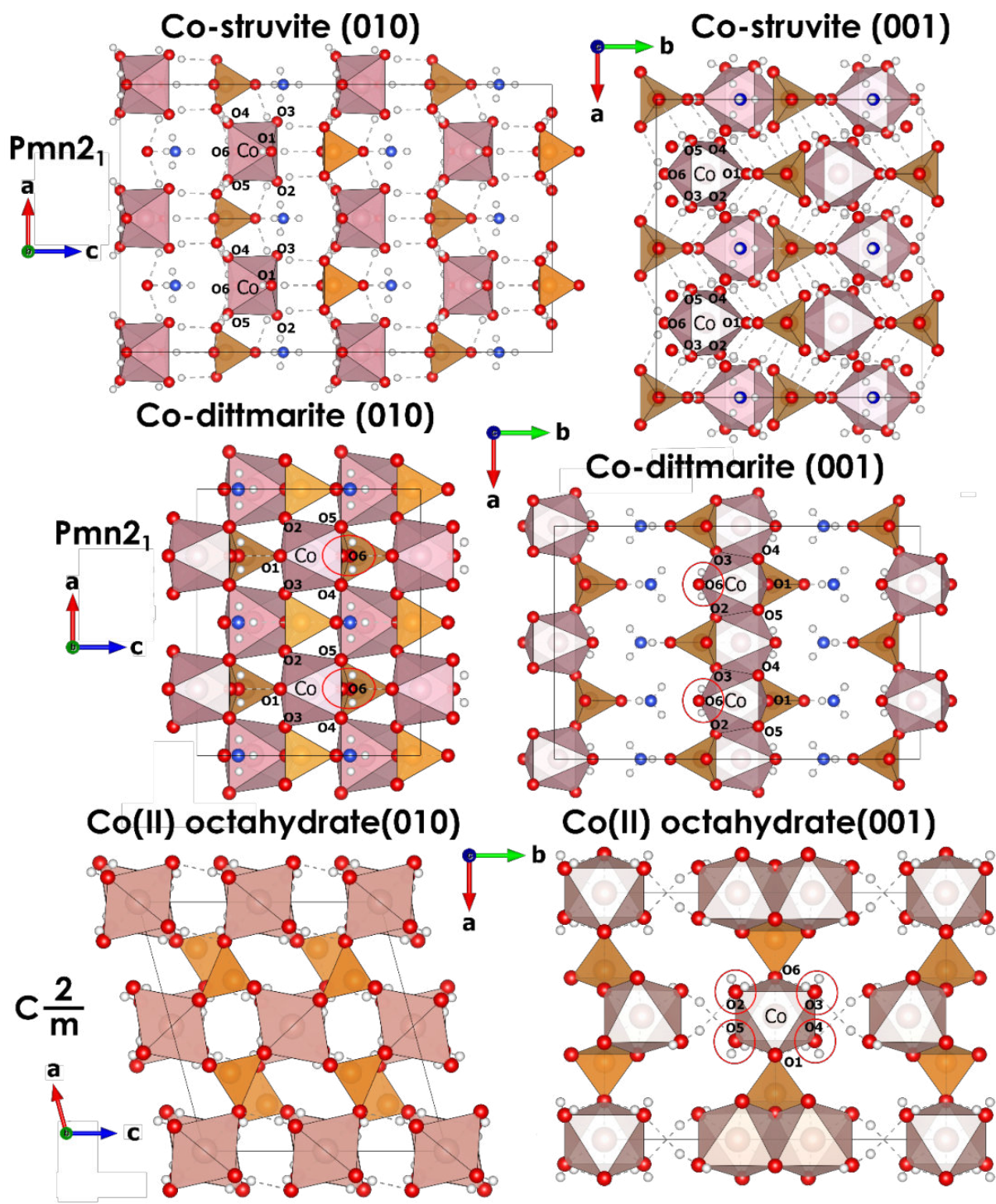

Figure S16: The (010) and (001) crystal planes of Co-struvite $NH_4CoPO_4 \cdot 6H_2O$, Co-dittmarite $NH_4CoPO_4 \cdot H_2O$ and Co(II)phosphate octahydrate $Co_3(PO_4)_2 \cdot 8H_2O$ crystal structure is displayed with the respective space group $Pmn2_1$ for Co-struvite and Co-dittmarite (orthorhombic) and space group $C2/m$ (monoclinic); Co: light pink, P: orange, O: red; N; blue and H: white; one or two exemplary $CoO_6$ octahedrons are marked by atom labels with all the coordinating oxygen numbered; the coordinating crystal water in Co-dittmarite and Co-octahydrate is marked with red ellipsoids; note that in the Co-dittmarite and Co-octahydrate structures, the $CoO_6$ is linked to the phosphate tetrahedrons, while in the Co-struvite structure the $CoO_6$ is isolated by six coordinating crystal water oxygens.

**Supplementary Note 1**

Cobalt is classified as a CRM according to the EU, because it is used in a number of technological applications, but it is only exploitable in few ore deposits in the world, mostly located in the Democratic Republic of the Congo. There, it is concentrated in minerals such as the cobaltite (CoAsS), erythrite ($Co_3(AsO_4)_2 \cdot 8H_2O$) or skutterudite ($CoAs_3$) in hydrothermal altered mantle rocks in association with mainly nickel. Although, it also plays an essential role in several enzymes and co-enzymes involved in the metabolism of animals and plants for example vitamin B12 (cobalamin), at above-trace concentrations cobalt is highly toxic.

Nickel is widely used in the industry and is also relatively common. It is mined in lateritic ultramafic rocks from ore minerals like pentlandite (($Ni,Fe)_8S_9$) or nickel ferrous limonite (($Ni,Fe)O(OH)$). Unfortunately, in the environment nickel compounds cause numerous carcinogenic effects in human and plants and it is one of the most common allergens in the world. This is even though at trace concentrations it also plays an important biological role for plants and microbes as a part of enzymes such as urease [10, 11]. Hence although nickel is not a CRM, it should be recovered from wastewater.

In mining waters of the Eshidiya mine (Jordan) the $PO_4^{3-}$ and $NH_4^+$ content ranges from 0.1 - 8.9 mg/l and 0.03 - 0.95 mg/l, respectively [5]. The related ground water in the same region showed concentrations values of $PO_4^{3-}$ and $NH_4^+$ at 3.5-3.8 and 0.01-0.05 mg/l. Mine wastewaters from the Duluth Gabbro complex in Minnesota (USA) contain average values of 198 mg/l of Ni and 9.8 mg/l of Co [4]. In the Katanga province (DRC) in effluent surface mine waters of a Co-Cu ore deposit Co reached maximum concentrations of 3.16 mg/l [1]. In Rio Piscianas (Italy) the highly contaminated groundwater from mining activities in the 1990s had concentrations of 1.5-2.9 mg/l for Co and 3.0-4.6 mg/l for Ni [2]. Ground water samples from the Amik plain (Turkey) contain in average 13 mg/l $NH_4^+$, 87 mg/l Phosphorus, 13 µg/l Co and 50 µg/l Ni [3].

These numbers indicate that transition metals as well as ammonia and phosphate are common contaminants in ground and waste waters (Table S3). As mining wastewaters contain high amounts of transition metals (mg/l) at low pH (~4-5) while agricultural waste- and ground waters demonstrate high concentrations of phosphate and ammonium at medium to high pH (7-9) a mixture of both wastewaters could lead to precipitation of transition metal phosphates in significant amounts. Due to the regularly presence of other metal cations in higher concentration in actual waste waters e.g. Mg or Ca, other phosphate phases will precipitate instead of M-struvite. Nevertheless, a P- and heavy metal recovery could be conceivable in synthetic, industrial waste waters, sludges, tailings where these cations may not be present. For example, waste waters derived as byproducts from recycling batteries, accumulators or catalysators could provide a limited composition where the precipitation of M-struvite could be possible in high amounts. Further experiments to the up scalability and implementation as synthetical wastewater treatment method would quantitively evaluate if the precipitation of M-struvite is applicable as P- and 3d metal recovery method or not. Additionally, a complete characterization of the pure transition metal struvite systems could be highly beneficial for adsorption/precipitation studies of 3d metals on common Mg-struvite [12-14].

**Supplementary Note 2**

The initial pH was ~7.9 for the 0.1 M DAP solution and ~5.7 for both $NiSO_4$ and $CoSO_4$ solutions at 0.1 M, and the 0.1 M $MgCl_2$ solution exhibited an initial pH of 7.0. When metal salt and DAP equimolar solutions were mixed, near-instant precipitation could be observed in both transition metal systems with the simultaneous drop in pH. On the other hand, the Mg-system had exhibited an induction time of ~5 s before it turned turbid. Immediately after mixing both 0.1 M solutions the pH dropped down to around ~5.8 for all systems. While the pH of the 0.1 M Mg and Ni sample declined slightly over the first 50 - 150 s with different slopes to a stable plateau value of 5.8, the pH for 0.1 M Co reached nearly instantaneously a plateau value of 5.9 and was only slightly until around 1500 s, marked by phase I (SI: Figure S4). After 1500 s a significant pH jump to a value of 4.6 over 500 s occurred in the 0.1 M Co sample marked by phases II and III which was most likely related to the transformation of the colloidal phases as the color changed simultaneously from purple to light pink.

The initial pH of the 0.02 M metal solutions was 6.5 (Mg), 5.9 (Ni and Co) for the low concentration runs. The same 0.1 M DAP solution was used as before. Similar trends as in the highly concentrated equimolar solution runs were observed except for Co although in all systems M-struvite formed. In this case, the pH of the Co mixture declined slowly after 450 s to 7.2 and stagnated at this value, as was the case for the other struvite samples. The PHREEQC equilibrium pH calculations are in a relatively close agreement with the experimental values within a confidence interval of ΔpH ≈ 0.3 (colored dotted lines in Figure S4). That indicates a progression of the precipitation reaction to completion for all the samples. The deviation between the calculated and measured pH trends, occurred probably due to the presence of amorphous phases, which were difficult to consider in the calculations as barely any (if any) thermodynamic data exist for them.

The Ni- and Mg-struvite formation reaction proceeds fast within the first 100 - 150 s, while the Co-phosphate reactions exhibit much longer reaction times (400 s for 0.02 M and around 3000 s for 0.1 M $Co^{2+}$) due to the preservation and transformation of long-lived colloidal phases.

**References SI**